\documentclass[nonblindrev]{informs3_modified}

\usepackage{graphics}
\OneAndAHalfSpacedXI

\usepackage{endnotes}
\usepackage{xcolor}
\usepackage{filecontents}
\usepackage{enumerate}
\usepackage{commath}
\usepackage{algorithm}
\usepackage[normalem]{ulem}
\usepackage{float}
\usepackage{algorithmic}
\usepackage{listings}
\let\footnote=\endnote

\definecolor{blue}{rgb}{0,0,0.9}
\definecolor{red}{rgb}{0.9,0,0}
\definecolor{green}{rgb}{0,0.9,0}

\def\cA{{\cal A}}
\def\cS{{\cal S}}

\newcommand{\R}{\mathbb R}

\newcommand{\M}{\mathbf M}
\newcommand{\V}{\mathbf V}

\newcommand{\I}{\mathbf I}

\newcommand{\EE}{\mathbb{E}}
\newcommand{\PP}{\mathbb{P}}

\newcommand{\cD}{\mathcal{D}}
\newcommand{\cH}{\mathcal{H}}
\newcommand{\cV}{\mathcal{V}}
\newcommand{\cE}{\mathcal{E}}

\newcommand{\cO}{\mathcal{O}}
\newcommand{\cU}{\mathcal{U}}
\newcommand{\cC}{\mathcal{C}}
\newcommand{\cN}{\mathcal{N}}
\newcommand{\cG}{\mathcal{G}}
\newcommand{\cT}{\mathcal{T}}
\newcommand{\ind}{\mathbf{1}}
\newcommand{\QED}{\ \hfill\rule[-2pt]{6pt}{12pt} \medskip}
\newcommand{\iN}{\cN^{in}}

\newcommand{\sgn}{\text{sgn}}
\newcommand{\sgplus}{f^+}

\newcommand{\eqd}{\mathop{=}\limits^d}

\usepackage{amsmath}
\usepackage{algorithm}
\usepackage{algorithmic}
\usepackage{appendix}

\def \P {\mathbb{P}}
\def \R {\mathbb{R}}

\def \ind {\mathbb{I}}

\usepackage{natbib}

\bibpunct[, ]{(}{)}{,}{a}{}{,}%
\def\bibfont{\small}%
\def\bibsep{\smallskipamount}%
\def\bibhang{24pt}%

\TheoremsNumberedThrough
\ECRepeatTheorems
\EquationsNumberedThrough

\begin{document}
\RUNAUTHOR{}
\RUNTITLE{}
\TITLE{Online Learning and Optimization Under a New Linear-Threshold Model with Negative Influence}

\ARTICLEAUTHORS{%
		\AUTHOR{Shuoguang Yang\thanks{Department of Industrial Engineering and Decision Analytics, The Hong Kong University of Science and Technology}, Shatian Wang\thanks{ Department of Industrial Engineering and Operations Research, Columbia University, New York, NY 10027}, Van-Anh Truong\thanks{ Department of Industrial Engineering and Operations Research, Columbia University, New York, NY 10027} } 
		\AFF{ \EMAIL{yangsg@ust.hk}, 
		 	\EMAIL{sw3219@columbia.edu}, \EMAIL{vt2196@columbia.edu}, 
	} 
}	

\ABSTRACT{\textbf{Problem definition:}  Corporate brands, grassroots activists, and ordinary citizens all routinely employ Word-of-mouth (WoM) diffusion to promote products and instigate social change. Our work models the formation and spread of negative attitudes via WoM on a social network represented by a directed graph. In an online learning setting, we examine how an agent could simultaneously learn diffusion parameters and choose sets of seed users to initiate diffusions and maximize positive influence. In contrast to edge-level feedback, in which an agent observes the relationship (edge) through which a user (node) is influenced, we more realistically assume \emph{node-level feedback}, where an agent only observes when a user is influenced and whether that influence is positive or negative. \textbf{Methodology/results:} We propose a new class of negativity-aware Linear Threshold Models.  We show that in these models, the expected positive influence spread is a monotone submodular function of the seed set. Therefore, when maximizing positive influence by selecting a seed set of fixed size, a greedy algorithm can guarantee a solution with a constant approximation ratio. For the online learning setting, we propose an algorithm that runs in epochs of growing lengths, each consisting of a fixed number of exploration rounds followed by an increasing number of exploitation rounds controlled by a hyperparameter. Under mild assumptions, we show that our algorithm achieves asymptotic expected average scaled regret that is inversely related to any fractional constant power of the number of rounds. \textbf{Managerial implications:} During seed selection, our negativity-aware models and algorithms allow WoM campaigns to discover and best account for characteristics of local users and propagated content. We also give the first algorithms with regret guarantees for influence maximization under node-level feedback.
}%

\maketitle

\section{Introduction}
As online social networks become increasingly integrated into our daily life, popular platforms such as Facebook, Twitter, and YouTube have turned into important and effective media for advertising products and spreading ideas. Commercially, it has become routine for brands to use the word-of-mouth effects to promote products on social networks. In other spheres, politicians, activists, and even ordinary people can leverage these networks to instigate political and social changes.
Given the immense power of social networks in spreading information and ideas, it is not uncommon to see social network marketing campaigns backfire. Even when a campaign is carefully designed, negative reactions might still arise due to the controversial nature of the information being propagated. Therefore, it is necessary to consider frameworks that allow the formation and spread of negative attitudes, the likelihood of which depends on heterogeneous demographics.
        
\textbf{Problem of interest.} Motivated by the potential emergence of negative attitudes in social networks, we consider \emph{a negativity-aware multi-round influence maximization problem}. In our problem, an agent, hoping to promote certain information, conducts a marketing campaign over a time horizon -- for example, three months. The time horizon is further divided into rounds, such as one-week periods. At the beginning of each round, the agent selects a fixed-cardinality \textit{seed set} of users, called \emph{influencers}, in the network. These users initiate a cascade of information spread through the network. The agent then closely monitors the subsequent influence diffusion process in the social network. The rounds are independent and the round rewards are cumulative. The agent is aware of the potential emergence of negative reactions and possible negative influence during the diffusion process, but is initially unaware of the underlying parameters that govern the attitude diffusion. Her goal is to simultaneously perform two actions: first, to learn the parameters via the feedback she gathers during monitoring; second, to select the seed set in each round in order to maximize the total expected number of positively influenced users over all rounds. 

Our problem is relevant to the (Online) Influence Maximization literature. While most existing works model only positive influence, the works that do consider the spread of negative attitude are either not flexible enough to capture important real-world characteristics or are intractable due to a lack of desirable mathematical properties. Also, to the best of our knowledge, there is no influence maximization framework that captures both online learning and the potential spread of negative attitudes.
        
\textbf{Overview of results.} In the \textbf{first part} of the paper, we propose a novel class of Linear Threshold models that incorporates the formation and spread of negative attitude. We call such models \emph{negativity-aware}. We show that in these models, the expected positive influence function is monotone submodular. Thus we can use a greedy algorithm to construct seed sets of fixed sizes with constant approximation guarantees, when the objective is to maximize expected positive influence. Our models are flexible enough to account for both the heterogeneity in features of local users, such as value judgements and relationship strengths, and in features of the information being propagated in the diffusion, such as the relative dominance of a certain attitude.
To the best of our knowledge, we are the first to propose linear-threshold-based negativity-aware diffusion models that have such flexibility as well as monotone submodular objectives.

In the \textbf{second part}, we analyze an \emph{online-learning} setting for a multi-round influence-maximization problem, where an agent is actively learning the diffusion parameters over time through \textit{node-level} feedback while trying to maximize total cumulative positive influence. We assume that in each diffusion step, the agent observes whether a node becomes positively or negatively influenced, or remains inactive. This \textit{node-level feedback} assumption reflects the reality that network activity is typically restricted to two measurable observations: first, that while we are able to identify an activated user, we are not able the observe the specific contributions of his neighbours; second, that we are able to observe the time of activation. For example, on Twitter, assume Charlie is a follower of both Andrew and Bob. If Andrew and Bob both retweet a story, and Charlie further retweets that story, we cannot determine whether Andrew's influence on Charlie was stronger than Bob's. However, if Andrew and Bob tweeted on Monday, Charlie tweeted on Tuesday, and David tweeted on Wednesday, we would know that David's tweet could not have influenced Charlie's.
        
        We exploit a novel property of greedy algorithms that we call \emph{stability}.  This property implies that as the estimated parameters become sufficiently accurate, the greedy solution behaves as if it was obtained under the exact parameters. We prove that stability holds almost surely during the execution of the algorithm.
        
        Using the property of stability, we develop online learning algorithms that achieve asymptotic expected average scaled regrets in the order of one over the number of rounds to the power of any constant smaller than one.
        These are the first regret guarantees for node-level feedback models for influence maximization of any kind. 
        Subsequent to our work, a paper by \cite{li2020online} studies online algorithms under vanilla LT models and proposes two online algorithms, LT-LinUCB and OIM-ETC. Nevertheless, the implementation of LT-LinUCB relies on an offline oracle that is only computational tractable for very limited graphs, such as Bipartite graphs with in-degrees at most 2; OIM-ETC, while using an idea similar to ours, requires the knowledge of an unknown constant to balance exploration-exploitation trade-off and cannot be generalized to the online learning setting of LT-N models. We defer the detailed discussion of these approaches to Section \ref{sec:literature}.

The rest of the paper is organized as follows: in Section \ref{sec:literature}, we give a review of the classical information-diffusion models and the influence maximization problem in the online-learning setting.  We also summarize existing works on negativity-aware variants of these models. In Section \ref{sec:LT-N}, we introduce our LT-based negativity-aware diffusion models and prove their monotonicity and submodularity properties. In Section \ref{sec:ON} and \ref{sec:ON-LTN}, we introduce an online-learning version of our problem.  We propose an online-learning algorithm and benchmark its performance against an algorithm that has access to the exact diffusion parameters. In Section \ref{NE}, we conduct numerical experiments to compare the performance of our proposed algorithm with several baseline algorithms on a Twitter subnetwork.

\section{Literature Review}\label{sec:literature}
Researchers have proposed various diffusion models for information spread and have extensively explored ways to maximize the spread of influence in these models. In their seminal work, \cite{kempe2003maximizing} proposed the so-called \emph{Influence Maximization (IM)} problem. In IM, a social network is modeled as a directed graph $\cG =(\cV, \cE)$ where each node $v$ in the node set $\cV$ represents a user and a directed edges $e = (u,v)$ in the edge set $\cE$ indicates that information can spread from user $u$ to $v$. They considered a viral marketing problem on the graph $\cG$, where a decision maker seeks to identify an optimal set of $K$ seed users to initiate an influence diffusion process, so that the expected number of people eventually influenced by information diffusion is maximized. They put forward two diffusion models, the \emph{Independent Cascade Model (IC)} and the \emph{Linear Threshold Model (LT)}, that have been used almost exclusively in the subsequent influence maximization literature.  We will describe these models briefly. 

In the IC model, each edge $e = (u,v)$ has an associated \textit{weight}, which is denoted as $w(e)$. This weight measures the likelihood with which user $u$ successfully influences user $v$. We use $w$ to represent a function from $\cE$ to $[0,1]$ that maps each edge to its corresponding weight. We refer to the function $w$ as \textit{weights}. IC specifies an influence diffusion process in discrete time steps. Initially, all nodes are \textit{inactive}. In step $0$, a seed set $S$ of users is selected and \textit{activated}. In each subsequent step $s$, each user activated in step $s-1$ has a single chance to activate her inactive \textit{downstream neighbors}, independently with success probabilities equal to the corresponding edge weights. This process terminates when no more users can be activated. The set of users activated during the IC process is precisely the set of users who have been influenced by the information.

The LT model, on the other hand, focuses more on describing the \emph{combined} effect of neighbors in influencing a node. In this model, each edge $e$ is still associated with a weight $w(e) \in [0,1]$. Again we use $w$ to denote a function from $\cE$ to $[0,1]$ that maps each edge to its corresponding weight and refer to the function $w$ as \textit{weights}. It is also assumed that the sum of the incoming edge weights for each node is at most one. That is, $\sum_{(u,v)\in \cE} w(u,v) \leq 1 \;\; \forall v \in \cV$. The LT diffusion process also unfolds in discrete time steps. In step $0$, all nodes in the seed set $S$ becomes activated, and each non-seed node $v \in \cV \backslash S$ independently samples a threshold $b_v\sim \cU[0,1]$, i.e., uniformly from $[0,1]$. In each subsequent step $s$, for each inactive node $v$, if $$\sum_{(u,v)\in \cE, u \text{ activated}} w(u,v) \geq b_v,$$ then $v$ becomes activated. This process terminates after step $s$ if no nodes change their activation status in this step.

Given a diffusion model, let $f_w(S)$ denote the expected number of nodes activated during the diffusion process given the seed set $S$ and diffusion parameters $w$. We say that $f_w(\cdot)$ is \emph{monotone} if for any $S \subset T \subset \cV$, $f_w(S) \leq f_w(T)$. If for any $S \subset T \subset \cV$ and $v \in \cV \setminus T$, $f_w(S \cup \{v\}) - f_w(S) \geq f_w(T \cup \{v\}) - f_w(T)$, then we say that $f_w(\cdot)$ is \emph{submodular}. 

\cite{kempe2003maximizing} showed that it is NP-hard to find $S \in \argmax_{S \subset \cV} f_w(S)$ with respect to either the IC or LT model, using reductions from the maximum coverage problem and the vertex cover problem, respectively. They also showed that $f_w(\cdot)$ is both monotone and submodular with respect to the two diffusion models. As a result, a greedy-based algorithm can find a seed set $S^g \subset \cV, |S^g| \leq K$ such that $f_w(S^g) \geq (1-1/e-\epsilon) \cdot \max_{|S| \leq K} f_w(S)$ for any $\epsilon>0$ \citep{Nemhauser1978}. The additional $\epsilon$ term is due to the error in function estimation, as $f_w(\cdot)$ is \#P-hard to compute \citep{Chen10}. 
On the inapproximability side, \cite{Feige} showed that the maximum coverage problem is NP-hard to approximate within factor $1-1/e+\epsilon$ for any $\epsilon > 0$. This result implies that the gap between the lower and upper bound on the approximation guarantee for IM under the IC model has been closed. No such inapproximability result had been known for IM under the LT model until \cite{undirected} considered IM specifically in the undirected graph setting. They showed that even for this special case, IM under either IC or LT is APX-hard (i.e., there exists a constant $\tau$ such that approximating IM to within factor $1-\tau$ is NP-hard).

Influence maximization has also been studied in the adaptive setting, where the decision maker chooses one seed user at a time, based on the realizations observed so far \citep{adaptive, adaptive2, adaptive3}. An interesting observation made by \cite{adaptive3} worth pointing out is that although adaptivity allows the decision maker to choose seeds with more information, it might make him act in a more myopic way, and result in worse performance.

In the next subsections, we focus on literature most closely related to our setting. Readers interested in developing a holistic view on the broad area of information diffusion in social networks could refer to the survey by \cite{survey}.

\subsection{Negativity-Aware Diffusion Models}\label{subsec_neg_aware_diff_models}
The existing models for influence diffusion primarily focus on the spread of one attitude of influence, which we can consider as a positive influence for simplicity. More precisely, whenever a user is influenced during the information diffusion process, she adopts a positive attitude towards the information being spread. However, in practice, we cannot guarantee such  uniformity in attitude, especially when the message being promoted is controversial in nature.

A few authors were motivated to consider potential negative reactions and the spread of negative attitudes \citep{chen2011influence, neg1, neg2, neg3, neg4}. They proposed new negativity-aware models that allow a node to become either positively or negatively influenced. In these models, the basic influence maximization problem is to identify a seed set of size $K$ that maximizes the number of \emph{positively} influenced nodes.


\cite{chen2011influence} proposed the first negativity-aware model.  In addition to the influence probabilities $w$, they assume that there is a \emph{quality factor} $q\in [0,1]$ representing the quality of the product being promoted. While the activation process follows that of IC, once a node is chosen as a seed node or is activated by a positive upstream neighbor, it becomes positive with probability $q$ and negative with probability $(1-q)$, independently of everything else.  Meanwhile, if the node is influenced by a negative upstream neighbor, it becomes negative with certainty. 
\cite{chen2011influence} showed that the expected final number of positively influenced nodes is monotone and submodular. 
Their model has a strong negativity bias.  Any node activated by a negative upstream neighbor can only turn negative. In reality however, when the information being propagated is controversial, a person might be influenced by her friends' strong attitudes to look into the issue, but can develop a different attitude towards it. Another limitation of this model is that $q$ cannot be a function of individual nodes, reflecting users' individual attitudes.  
In Appendix Section \ref{apd:IC-N-break}, we provide an example in which the greedy algorithm can have an arbitrarily bad approximation ratio when the quality factors are heterogeneous.


\cite{neg1}'s model is richer as there are now four different types of users, i.e., (dis)satisfied and (non-)complainers. Each type has a different but fixed probability of participating in negative word-of-mouth. However, the model is still not flexible enough to account for the richness of user characteristics. Other more refined models are generally intractable \citep{neg2,neg4, neg3}, either being NP-hard to approximate within constant approximation guarantees, or being too involved for theoretical analysis.

\noindent \textbf{Competitive contagion} is an increasingly popular variant of IM that is similar to negativity-aware diffusion to some extent. In this variant, multiple agents are promoting competing products in the same social network through seeding influence diffusion. Most related works focused on developing strategies for the last mover given the seed sets chosen by all competitors, either to maximize its own influence spread or to minimize the influence spread of the competitors.
One could argue that the relevant models could also be used to study negativity-aware diffusion when one agent spreads positive information while the other spreads negative information.

Most existing models for competitive contagion assume tie-breaking rules that are realistic for the spread of substitute products from competing companies but not much so for the diffusion of positive or negative attitudes. For example, \cite{He11, Budak11, Zuo20, Kahr20} all assume that one pre-specified attitude dominates when a node is simultaneously exposed to more than one attitude. While some extent of dominance of a certain attitude is possible, we deem it too strong to assume such complete dominance in the negative-aware setting. In our model, we incorporate each individual's independent value judgments, which could be used to model a more flexible node-level dominance of attitude. 
\cite{Bharathi07}, \cite{Carnes07}, and \cite{Goyal19} proposed competitive contagion models that allow tie breaking to be random and dependent on the relative number of active neighbors with each attitude. However, the randomness does not take into account the strength of relationships: a closer friend's attitude could be weighted more heavily than that of an acquaintance. \cite{Borodin10}'s extended LT model is the closest to ours. It also postulates that both the positive and the negative attitude from friends jointly make one aware of a piece of information; the probability that an individual adopts the positive attitude is a weighted fraction of his active friends weighted by the respective edge weights. This model, however, is neither monotone nor submodular. It also does not take into account the relative dominance of a certain attitude or individuals' independent judgments.

To the best of our knowledge, we are the first to propose negativity-aware diffusion models that are not only flexible enough to incorporate a variety of individual user characteristics but also have monotone submodular objective functions. We allow users with different characteristics to have different information-sharing behaviors and attitude-formation patterns. Due to the monotonicity and submodularity of the objective functions, we can use a greedy algorithm to obtain a $1-1/e - \epsilon$-approximate solution, where $\epsilon$ is an error term that is caused by the (typical) use of simulation to evaluate the influence function \citep{Chen10}.

\subsection{Online Learning for Influence Maximization}
There is another line of work that focuses on the online-learning setting for influence maximization under the IC model \citep{OIM, IMB, CUCB, contextualIM, DISBIM, wen2017online}. In this setting, an agent starts with zero knowledge of the edge weights, and has $T$ rounds to advertise a product. In each round, it can select a seed set of up to $K$ nodes based on information observed in previous rounds, called \textit{feedback}.  The goal is to maximize the total expected influence spread over all rounds.

Two feedback mechanisms have been proposed. Under the \textit{edge-semi-bandit feedback}, the agent observes whether each \textit{activated} node's attempts to activate its downstream neighbors succeeded or not. On the other hand, under the \textit{node-level feedback}, the agent observes only the identity of the newly activated nodes in each diffusion step. More precisely, when a node $v$ is activated in step $s$ and more than one of its upstream neighbors were activated in step $s-1$, it is not possible to discern which of these upstream neighbors activated $v$. For IC-based diffusions, both node-level feedback and edge-level feedback can be assumed. LT-based models, on the other hand, assume the joint effort of active parent nodes in activating a given child node. As a result, for LT-based diffusions, the node-level feedback mechanism should be the only natural setup.

To date, the edge-semi bandit feedback setting has been well-characterized by various authors \citep{CUCB, wen2017online}, but not the node-level feedback setting.  \cite{IMB} used Maximum Likelihood Estimation(MLE)-based techniques to learn from node-level feedback, but do not provide regret guarantees for their MLE-based learning algorithm.

The difficulty of node-level learning lies in the special observation structure. When a set of edges are observed simultaneously, the agent is only able to estimate the combined effect of these edges but not the mean on every single edge. In this case, the upper bound constructed in UCB-type algorithms might not improve. \cite{li2020online}, in a paper subsequent to our work, assume access to a strong oracle: let $\cC$ be the current confidence region in the space of all possible edge weights $\mathcal{W}$ and $w \in \mathcal{W}$ be the true edge weights,
the oracle is able to find a seed set $\tilde S$ and a corresponding edge weights function $\tilde w \in \cC$ such that $|\tilde S| = K$ and with probability at least $\gamma$, $$f_{\tilde{w}}(\tilde S) \geq \alpha \cdot \max_{ w' \in \mathcal{C}}\max_{S\subseteq V, |S|=K} f_{w'}(S)$$ for some $\alpha$. Such an oracle enables them to prove a $\tilde \cO(
 \sqrt{T})$ $\alpha\gamma$-scaled regret. However, the implementation of such an oracle is computationally expensive for general networks. In particular, their proposed method first discretizes a continuous ellipsoid ($\cC$) by $\epsilon$-Net covers, and then uses the greedy oracle to find an $(1-1/e)$-approximation solution with respect to one edge weights function in each length-$\epsilon$ set. The total number of times the greedy oracle is invoked in this method is $\Theta((1/\epsilon)^{|\cC|})$ with $|\cC|$ being the dimension of $\mathcal W$. The authors also provide an Explore-then-Commit (ETC) algorithm which uses similar ideas to  our exploration-exploitation approach. However, to compute the number of required exploration rounds, one needs to have access to a parameter $\Delta_{\min}$: let $S^{\ast}$ be an optimal size $K$ seed set with respect to the true edge weights $w$, define $\cS_B := \{ S: f_w(S) < \alpha \cdot f_w(S^{\ast} )  \}$ to be the ``bad'' sets that do not achieve an $\alpha$-approximation, 
 \begin{equation*}
    \begin{split}
        \Delta_{\min}:= & \alpha \cdot  f_w(S^*) - \max_{S\in\cS_B} \{ f_w(S)\}.
    \end{split}
\end{equation*}
Our online learning algorithms, on the other hand, do not rely on such a strong oracle nor any prior knowledge of $\Delta_{\min}$. Furthermore, while our algorithm and analysis generalize to the online learning setting of LT-N models, their main result is not generalizable to the negativity-aware setting.

\noindent \textbf{Offline learning} A parallel line of work focuses on learning either the network structures or the influence functions from diffusion traces. For instance, \cite{TraceComplexity} analyzed the number of distinct traces, or node activation events with timestamps, that are required to reconstruct the network structure with high fidelity. Traces are similar to node-level feedback, both suffering from the uncertainty of which active nodes caused later activations. \cite{TraceComplexity} showed that for perfect inference on general connected graphs, no useful information could be extracted from the tail of a trace that consists of nodes activated after the first two. As will be detailed later, our online learning algorithm also primarily uses the front portion of each observed diffusion to learn edge weights.

\cite{Narasimhan15} established PAC learnability of influence functions ($f: 2^V \rightarrow [0,1]^V$ that maps each seed set to a vector specifying the activation probability of every node in the graph) for LT, IC, and the Votor model. 
While PAC learnability of influence functions is an important topic, it is quite orthogonal to online influence maximization. In the former, seed sets in the samples are assumed to be drawn i.i.d according to some fixed distribution and the goal is to learn the influence function itself; in the latter, the seed sets are decisions made by the learning agent whose primary objective is to maximize the cumulative influence spread.

\section{Negativity-Aware Diffusion Model}\label{sec:LT-N} 
In this section, we introduce a new negativity-aware diffusion model based on the Linear Threshold model, which we refer to as the Negativity-Aware Linear Threshold (LT-N) model.

In LT-N, each node can be in one of the three possible states at any time: \textit{positive}, \textit{negative}, and \textit{inactive}. Positive (resp. negative) means that the node holds a positive (resp. negative) attitude towards the information being propagated. Meanwhile, inactivate means the node has not yet developed any attitude towards the information, due to, for example, lack of awareness. Let $\sgn(v) = +1, -1, 0$ denote $v$ being positive, negative, or inactive, respectively. We assume that,  initially, all nodes are in the inactive state. In other words, $\sgn(v)=0$ for all $v$.

A person's attitude is not only determined by her friends' but also by her own experience and value judgment. To incorporate such personal bias, we introduce two \textit{autonomy factors} associated with each node  $v \in \cV$, $q^+(v), q^-(v) \geq 0$ such that $q^+(v) + q^-(v) \leq 1$. The autonomy factors for each node depend on the information being promoted, as well as on the node's unique characteristics. The bigger $q^+(v)$ is relative to $q^-(v)$, the more $v$ leans towards the positive attitude. For now, we assume that $q^+$ and $q^-$ are both known.  

For each node $v$, we refer to the sum of its autonomy factors as its \emph{belief score} which is denoted by $r(v)$, i.e.,
\[r(v) = q^+(v) + q^-(v).\]
It measures the amount of trust that the node places in her own judgment. Intuitively, the smaller $r(v)$ is, the more susceptible $v$ is to others' attitudes. 

A person also tends to place different weights on different friends' influences.  We model this by having a weight $w(e) \geq 0$ associated with each edge $e = (u,v) \in \cE$.  The larger $w(e)$ is, the more influential $u$'s attitudes is on $v$. We assume that for each node $v \in \cV$, the sum of weights of its incoming edges lies between 0 and 1. More precisely, let $\iN(v) = \{u: (u, v) \in \cE\}$ be the set of in-neighbors of $v$, we assume that
\[\sum_{u \in \iN(v)} w(u, v) \in [0,1],\ \forall v \in \cV.\]

During the LT-N diffusion process, we assume that positive and negative influences from friends, rather than canceling each other out, \emph{jointly} prompt each person to take note of the information being diffused. Intuitively, the fact that a piece of information triggers different reactions by people around us should further pique our interests to learn about it, and eventually to develop our own attitude towards it. Subsequently, in our model, a node is \emph{activated} the first time the sum of weights from its active neighbors exceeds a threshold. After being activated, the node decides on its attitude (positive or negative) based on the ratio between the positive influence (sum of weights from positively activated friends) and negative influence (sum of weights from negatively activated friends) that are exerted on the node most recently, as well as the node's autonomy factors. 

Mathematically, the LT-N diffusion process unfolds in discrete time steps as follows, starting from seed nodes in the chosen seed set $S$. (We reserve ``round'' for online learning). 
\begin{itemize}
    \item Each node $v \in \cV \backslash S$ independently samples a threshold $b_v\sim \cU[0,1]$.
    \item In step $0$, all nodes are inactive. Set $A_0 = A_0^+ = A_0^- = \emptyset$.
    \item In step $1$, all seed nodes become positive, and all non-seed nodes $u \in \cV \backslash S$ are inactive. Set $A_1=A_1^+=S$ and $A_1^-=\emptyset$.
    \item In general, let $A_{\tau}$ (resp. $A^+_{\tau}$, $A^-_{\tau}$) denote the set of nodes that are activated (resp. positive, negative) by the end of time step $\tau\geq 0$. In each subsequent time step $\tau = 2, ...$, for each inactive node $v\in \cV \setminus A_{\tau -1} $, if
    \[\sum_{u \in \iN(v) \cap A_{\tau-1}} w(u,v) \geq b_v,\] 
    then $v$ becomes active. It turns positive with probability  
    \begin{align}\label{p_Atau+_cond_Atau}
    \PP[v\in A_{\tau}^+ | v\in A_{\tau}] =r(v) \cdot \frac{q^+(v)}{q^+(v)+q^-(v)} + [1-r(v)] \frac{\sum_{u \in \iN(v) \cap (A^+_{\tau-1}\setminus A_{\tau-2})} w(u, v)}{\sum_{u \in \iN(v) \cap (A_{\tau-1}\setminus A_{\tau-2})} w(u, v)},
    \end{align}
    and negative otherwise. Note that the probability of the node turning positive or negative is a convex combination of its own belief and the most recent influences from its active neighbours.
    \item The process terminates when no more inactive nodes can be activated. Let $A(S)$ (resp. $A^+(S)$, $A^-(S)$) denote the set of active (resp. positive, negative) nodes at the end of the process, which runs until (at most) $\tau = |\cV \setminus S|$.
\end{itemize}

 Note that a node must become either positive or negative once activated. Meanwhile, as in the original LT diffusion model, the nodes that are activated in the current time step $\tau$ does not affect other nodes in the same time step.

\subsection{Influence Maximization}\label{sec:mono+subm}

Under the LT-N model, we consider the problem of choosing at most $K$ seed nodes to maximize  the number of positive nodes at the end of the diffusion process. More rigorously, let $\sgplus(S)$ be the expected number of positive nodes at the end of the diffusion process under LT-N. Our goal is to maximize $f^+(S)$, subject to the cardinality constraint that $|S| \leq K$ for some positive integer $K$.

This problem is an extension of the influence maximization under the original Linear Threshold model, which is NP-hard \citep{kempe2003maximizing}. In Theorem \ref{thm:sig^+_mono_subm} below, we prove that $f^+(S)$ is monotone submodular under LT-N. \cite{Nemhauser1978} showed that when the set function one wants to maximize is monotone and submodular, then the greedy algorithm guarantees a $(1-1/e - \epsilon)$-approximation. Therefore, it follows that greedy is a $(1-1/e-\epsilon)$-approximation algorithm for our problem.
\begin{theorem}\label{thm:sig^+_mono_subm}Let $\sgplus(S)$ be the expected number of positive nodes at the end of the diffusion process under LT-N given seed set $S$. Then, $\sgplus(\cdot)$ is monotone submodular.
    \end{theorem}
    \proof{Proof sketch.}
    We define another diffusion model that we call the \emph{negativity-aware triggering set model (TS-N)}. We first show that the expected positive influence spread function of TS-N is monotone submodular. We then show that the set of positively (negatively) activated nodes in each step of LT-N has the same distribution as that in TS-N. This way, we conclude that the expected positive influence function of LT-N is also monotone submodular. Our result is a generalization of Theorem 4.6 in \cite{kempe2003maximizing}. In our proof, we use a new ``correction-sampling'' together with the Multinomial in-neighbor sampling to construct TS-N so that it models nodes' attitudes as well. The details of the proof are provided in Appendix Section \ref{sec:LT-N-mon-sub}.\QED

\section{Learning From Node-Level Feedback} \label{sec:ON}

The previous two sections are based on the assumption that both edge weights $w: \cE \to [0,1]$ and autonomy factors $q^+,q^-: \cV \to [0,1]$ are known. In this section we consider an \emph{online learning} setting of this problem. Namely, the autonomy factors and the edge weights are initially unknown and need to be gradually learned. We further assume that the autonomy factors and the edge weights can be linearly generalized. More specifically, we assume that there exist two \textit{unknown} vectors $\theta^* \in \mathbb{R}^d$ and $\beta^* \in \mathbb{R}^{d'}$. For each $e \in \cE$, we have a \textit{known} feature vector of edge $e$, $x(e) \in \mathbb{R}^{d}$, such that $w(e) = x(e)^\top \theta^*$. For each $v \in \cV$, we have two known feature vectors $x_+(v), x_-(v) \in \mathbb{R}^{d'}$ such that $q^+(v) = x^{\top}_+(v)\beta^*, q^-(v) = x^{\top}_-(v)\beta^*$. With the linear generalizations, learning the autonomy factors and the weights amounts to learning the corresponding unknown vectors $\theta^*$ and $\beta^*$.

Recall that the node activation process in our LT-N model follows the classical LT model. After a node $v$ is activated, the sign of the activation (positive or negative) depends on both the autonomy factors $q^+(v), q^-(v)$ and the most recent influences from its active friends, as defined in \eqref{p_Atau+_cond_Atau}:
\begin{align*}
    \PP[v\in A_{\tau}^+ | v\in A_{\tau}] =r(v) \cdot \frac{q^+(v)}{r(v)} + [1-r(v)] \frac{\sum_{u \in \iN(v) \cap (A^+_{\tau-1}\setminus A_{\tau-2})} w(u, v)}{\sum_{u \in \iN(v) \cap (A_{\tau-1}\setminus A_{\tau-2})} w(u, v)},
\end{align*}
where $r(v) = q^+(v) + q^-(v)$ is the belief score of $v$ defined previously.

Our plan is to learn $\theta^*$ for the weight function $w:\cE \to [0,1]$ from the node activation observations. For this part, we do not use the observed signs of the activation. Therefore, our result is suitable for the online learning setting with respect to the classical LT model. For this reason, we present the learning framework under the classical LT models in this section.    As for learning $\beta^*$ for the autonomy factors $q^+, q^-: \cV \to [0,1]$ with respect to LT-N, we use the signs of the observed node activation. In the next section, we extend the framework to account for the learning of $\beta^*$.

Consider the classical LT models, in each round, the agent activates a seed set that initiates the information diffusion on the network. Unlike the \emph{edge-level feedback} model (which is assumed by most existing online influence maximization literature under IC as the underlying diffusion model, where the status of each edge that takes part in the diffusion can be observed), our node-level feedback model assumes that the agent can only observe node status. More specifically, in each diffusion time step, the agent observes whether or not a node becomes positively or negatively activated or remains inactive, but she does not get to observe how each of its active parents contributes to this node's activation. Since several edges may contribute to a node's activation simultaneously, it is hard to discriminate the contribution of each edge and estimate the edge weights accurately. Because of this difficulty, online learning with node-level feedback has remained largely unexplored until this work. 

In this section, we first mathematically formulate the online learning problem and discuss some assumptions we impose. Then we investigate the key obstacles in learning and propose an algorithm that performs weight estimation and selects seed sets in each round. Finally, we conduct a theoretical analysis of the performance of the algorithm. Specifically, we show that the asymptotic expected average scaled regret of Algorithm \ref{alg:IM02} with hyper-parameter $q \in \mathbb{Z}_+$ is bounded by $\cO(T^{-q/(q+1)})$, where $T$ is the total number of rounds.

\subsection{Learning in Classical LT Model} 
\label{sec:ON-classical}
\cite{wen2017online} have investigated the performance of edge-level feedback IC model and purposed a UCB-type learning algorithm IMLinUCB. However, it is hard to extend their work to any node-level feedback model such as LT-N. The main challenge comes from parameter estimation. In round $t$, IMLinUCB estimates $\theta^*$ using ridge regression with realizations of independent Bernoulli trials on the edges observed so far. Denote the ridge regression estimate in round $t$ by $\theta_t$. IMLinUCB constructs a confidence ball around $\theta_t$, and derives the upper confidence bound (UCB) weight $U_t(e)$ for every edge $e \in \cE$. IMLinUCB then selects the seed set by feeding $U_t(e)$, $e\in\mathcal{E}$ to a greedy approximation oracle.

Intuitively, with more observations, the upper bound $U_t(e)$ converges to the true weight $w(e)$ on each edge $e\in \cE$, thus making the selected seed set an $\alpha$-approximation solution to the optimum.

\begin{figure}
    \centering
    \includegraphics[scale=0.7]{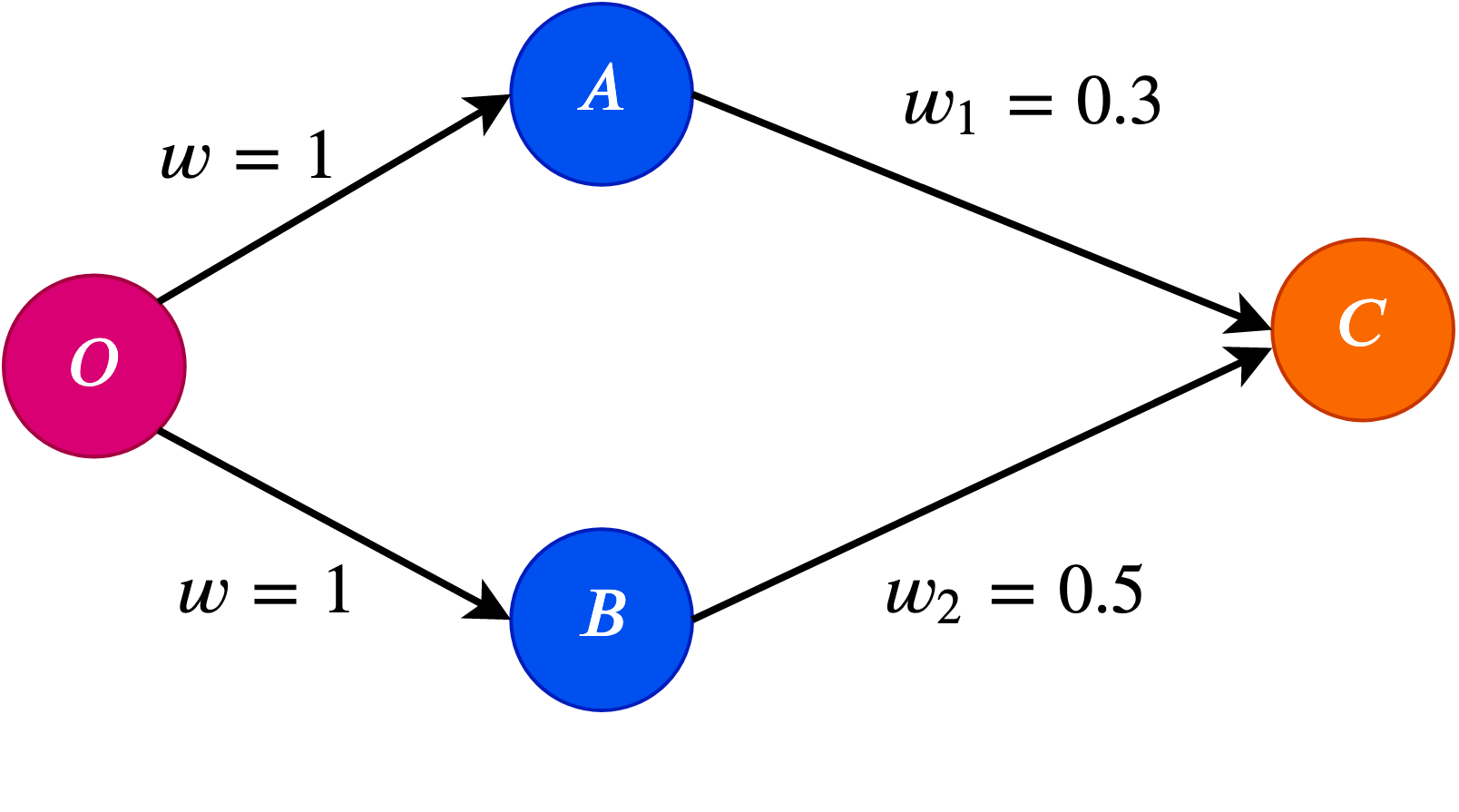}
    \caption{The case where weights and upper bounds cannot be estimated accurately by node-level observations.}
    \label{fig:2}
\end{figure}

Unfortunately, the structural similarities between their IM problem and the classical linear contextual bandit problem, and between the IMLinUCB and the classical algorithms in \cite{abbasi2011improved} do not hold anymore under node-level feedback. As several edges can simultaneously contribute to the activation of a single node, it is generally not possible to estimate the weight and upper bound on each individual edge accurately. Consider a simple example as illustrated in Figure \ref{fig:2}. We have two edges $e_1 = (A, C)$ and $e_2 = (B, C)$ with corresponding features $x(e_1) = (1,0)$ and  $x(e_2) = (0,1)$. Suppose that the true weights on $e_1$ and $e_2$ are $w_1 = 0.3$ and $w_2 = 0.5$ and these two edges are always observed simultaneously. In this case, with more observations, the estimation of $w_1 + w_2$ converges to $0.8$. However, if we try to estimate $e_1$ and $e_2$ separately, as these two edges are always observed together, we have $U(e_1) = 0.8$ and $U(e_2) = 0.8$. This example shows that the upper confidence bound of each individual edges does not necessarily converge to its true weight even if we have infinitely many observations of this edge.

The example above indicates that there is no quick extension of IMLinUCB for LT-N. Thus, further assumptions, as well as more sophisticated algorithms, are required to ensure an increasingly accurate edge weight estimation
as more node-level realizations are observed.
\subsubsection{Technical assumptions}\label{assumptions}
Recall that in each round $t$, the nodes are activated in discrete time steps according to our LT-N model, with nodes in the seed set $S_t$ being activated in time step $0$ of round $t$. For each node $v \in \cV$, we use $\tau_t(v)$ to denote the time step at which node $v$ becomes activated in round $t$. When $v\in S_t$, we have that $\tau_t(v)=0$. If $v$ is not activated in round $t$, then we set $\tau_t(v) = \infty$. 


For each node $v\in \cV \setminus S_t$, define its \emph{relevant parents} as follows: 
\begin{align*}\label{A(v)_activators}
    RP_t(v) := \begin{cases}
    \{ u \in \iN(v): \tau_t(u) \leq \tau_t(v)-1 \} & \text{if $\tau_t(v)<\infty$},\\
    \{ u\in \iN(v): \tau_t(u)<\infty \} & \text{if $\tau_t(v)=\infty$}.
    \end{cases}
\end{align*}
That is, the set of relevant parents $RP_t(v)$ is the set of nodes that are relevant to the activation status of $v$ in round $t$. We say the weight $w(e)$ on an edge $e=(u,v)$ is active if $u$ has been activated. When $\tau_t(v) < \infty$, $RP_t(v)$ is the set of its parent nodes who collectively push the sum of active incoming weights at $v$ to exceed $v$'s threshold for the first time. When $\tau_t(v) = \infty$, $RP_t(v)$ is the set of nodes that have collectively failed to push the sum of active incoming weights at $v$ to exceed its threshold. Note that for an inactive node $v$ in round $t$, $RP_t(v)$ might not be empty, since some of its parent nodes might be activated during the diffusion process but have failed to activate $v$. 

Our analysis is based on a few assumptions on the weights and solution stability, which we will state and justify below. Note that all norms used in the rest of the paper are $\ell_2$-norm, and are denoted by $\|\cdot\|$.

We first introduce  assumptions on the edge weights. 
The first assumption is the linear generalization of edge weights introduced previously. We assume that each edge $e \in \cE$ has an edge \emph{feature vector} $x(e) \in \R^d$ that characterizes the relationship between $e$'s two end nodes. The weight on each edge is a linear mapping from its feature. More formally, we have
\begin{assumption}[Linear parameterization]\label{assump:linear_param}
    There exists $\theta^*\in \R^d$, $\|\theta^*\|\leq D$ such that the true edge weights are $w_{\theta^*}(e) = x(e)^\top \theta^*\in [0,1]$. By the assumption on the incoming weight sum of our LT model, we have $\sum_{u\in \iN(v)} x(e)^\top \theta^* \leq 1$ for all $v\in \cV$.
\end{assumption}
Such a linear generalization of diffusion parameters is also used in \cite{wen2017online}.  The generalization makes our learning model more scalable. 
We use $\theta$ to denote a generic vector in $\R^d$ and refer to it as the \emph{parameter}. We refer to $\theta^\ast$ as the \emph{true parameter}. 

Furthermore, we assume that the ``aggregated'' features are bounded too:
\begin{assumption}[Feature regularity]\label{assump:upepr_bound_sum_features}
For all $v\in \cV$ and all $B\subseteq \iN(v)$, $\| \sum_{u\in B} x(u,v)\| \leq 1$.
\end{assumption}
Note that Assumption \ref{assump:upepr_bound_sum_features} is similar to the feature boundedness assumption in many existing works on contextual linear bandit problems. For example, \cite{wen2017online} assumes that the norms of the edge features are bounded. Similar assumptions are also made by \cite{abbasi2011improved} and \cite{chu2011contextual}. In addition, the LT-N model, like any other LT model, requires the sum of weights of incoming edges of every node to be bounded by $1$. It is thus natural to assume that the norm of the sum of any subset of incoming features at every node is bounded by 1, which can always be achieved by an appropriate scaling of the feature space.

One of our key ideas is to make sure that the features of observed edges are diverse enough to allow enough information to be collected in all directions of $\theta_t$,  so that $\theta_t \rightarrow \theta^\ast $ as $t \rightarrow \infty$. More specifically, we impose a feature diversity assumption as follows:
\begin{assumption}[Feature diversity] \label{assump:exploration_nodes}
    There exists $d$ edges $e^{\circ}_i$, $1\leq i\leq d$ such that the matrix $\sum_{i=1}^d x(e^\circ_i)x(e^\circ_i)^\top$ is positive definite. In other words, the minimum eigenvalue $\lambda_{\min}^\circ := \lambda_{\min}\left(\sum_{i=1}^d x(e^\circ_i)x(e^\circ_i)^\top\right)$ is strictly greater than $0$.
\end{assumption}
It is easy to see that the existence of $d$ edges with linearly independent features would be sufficient to ensure Assumption \ref{assump:exploration_nodes}.  This should be easy to satisfy as the dimension of the features is usually much smaller than the total number of edges, that is, $d \ll |\cE|$. Under this assumption, if we keep exploring those edges, the confidence region will shrink in all feature directions, so that $
\theta_t \rightarrow \theta^\ast$ as $t \rightarrow \infty$. 

Let $K$ be the maximum seed set cardinality. Let $f(S,w)$ be the expected influence spread given seed set $S$ and edge weights $w$. Note that previously we used $f_w(S)$ to denote the same function when the true edge weights function is known; In the online learning part, we use $f(S,w)$ instead to further emphasize that the expected influence spread is a function of both the seed set $S$ and the edge weights $w$. Denote $S^{\text{opt}}(w) = \arg\max \{ f(S,w): |S|\leq K,\ S\subseteq \cV \}$ and $f^{\text{opt}}(w) = \max \{ f(S,w): |S|\leq K,\ S\subseteq \cV \}$. 
As discussed in Section \ref{sec:mono+subm}, $f(\cdot,w)$ is monotone and submodular for any fixed edge weights $w$. Therefore, a greedy algorithm with exact evaluation of $f$ returns a $(1-1/e)$-approximation solution. Since evaluating $f$ is $\#$-P hard, we assume access to an approximation oracle: 
\begin{assumption}[Approximation oracle]\label{assump:alpha_beta_oracle} Let $\epsilon >0$ and $\alpha = 1-1/e - \epsilon$,
    there exists an efficient, possibly random $(\alpha,\gamma)$-\emph{oracle} that takes $\cG$, $w$, $K$, and outputs a seed set $\tilde S$ such that $f(\tilde S,w)\geq \alpha \cdot f^{\text{opt}}(w)$ with probability at least $\gamma$.
\end{assumption}
An example of $\alpha$ is $\alpha = 1-1/e -0.01$.  The reverse reachable set method by \cite{tang2014influence} can be easily extended to obtain such an approximation oracle. 

We also impose assumptions on the solution stability of the network. Consider the solution returned by the approximation oracle around $\theta^\ast$.
Define the \textit{$\alpha$-approximation seed sets with respect to $\theta \in \mathbb{R}^d$ as} $\cA(\theta,\alpha) = \{ S\subseteq\cV: f(S,w_\theta) \geq  \alpha\cdot f^{\text{opt}}(w_\theta)\}$. That is, $\cA(\theta,\alpha)$ is the set of seed sets whose expected influences with respect to edge weights $w_\theta(e) = x(e)^\top \theta$ are at least $\alpha$ times that of an optimal seed set. Our next assumption states that the set of  $\alpha$-approximation sets and the optimal seed set are invariant under small perturbations.
\begin{assumption}[Stability]\label{assump:invariant_Approx_seed_set} 
    There exists a constant $\epsilon>0$ such that for any $\theta \in \mathbb{R}^d$ that satisfies $\| \theta-\theta^*\|\leq \epsilon$, we have $\cA(\theta,\alpha) = \cA(\theta^*, \alpha)$. Moreover, there exists a seed set $S^{\cA}$ such that $S^{\cA} \in \argmin\{f(S,w_{\theta}): S\in \cA(\theta,\alpha)\}$ for all $\theta$ such that $\|\theta - \theta^*\| \leq \epsilon$. 
\end{assumption}
The above assumption will be satisfied under mild conditions. In Lemma \ref{thm:2} below, we provide a sufficient condition for Assumption \ref{assump:invariant_Approx_seed_set} to hold, and show that this sufficient condition holds with probability one.
\begin{lemma} \label{thm:2}
Let $\theta^*$ be the true parameter, $\cA(\theta^*,\alpha)$ be the set of $\alpha$-approximation sets with respect to $\theta^*$, and $f^{\text{opt}}(w_{\theta^*})$ be the optimal value with respect to $\theta^*$. Assumption \ref{assump:invariant_Approx_seed_set} holds whenever $\min_{S \in \cA(\theta^*,\alpha)} f(S,w_{\theta^*})> \alpha\cdot f^{\text{opt}}(w_{\theta^*})$.  This sufficient condition holds with probability 1 if we sample $\alpha$ uniformly from an interval $[1-1/e-\epsilon_1, 1-1/e-\epsilon_2]$ with $\epsilon_1 > \epsilon_2$.
\end{lemma}

The proof of Lemma \ref{thm:2} is in Appendix Section \ref{apd:online-lemmas-thm:2}. It shows that Assumption \ref{assump:invariant_Approx_seed_set} fails only when there is a set that provides exactly an $\alpha$-approximation, which happens with probability zero. 

The stability assumption is crucial for analyzing the theoretical performance of our algorithm. Suppose the current estimator $\theta$ is close enough to $\theta^\ast$ such that $\| \theta - \theta^\ast\| \leq \epsilon$, and the greedy algorithm returns a size-$K$ $\alpha$-approximation solution $S$ such that $f(S,w_{\theta}) \geq \alpha\cdot f^{\text{opt}}(w_{\theta})$. By Assumption \ref{assump:invariant_Approx_seed_set}, we have $S \in \cA(\theta,\alpha) = \cA(\theta^*, \alpha)$, which implies $S$ is also an $\alpha$-approximate solution to $\theta^\ast$, and it yields zero  regret. Although Assumption \ref{assump:invariant_Approx_seed_set} is general enough and provides a possibility of getting theoretical guarantees for online-learning, it has not been exploited by any work prior to ours, to the best of our knowledge. A subsequent work by \cite{li2020online} employs a stability idea similar to ours in the design and analysis of an algorithm called Exploit-then-Commit (ETC), without the linear generalization of edge weights. A core idea they used is that $|f(S, w) - f(S, w')|$ could be upper bounded by the sum of edge weight differences. While this result helps them to calculate how accurate the edge weights estimates need to be to obtain zero regret in the exploitation rounds, the idea is not generalizable to the LT-N setting, where the probability of an activated node turning positive depends on the relative size of the most recent positive influence the node receives.


\subsubsection{Performance metrics}
One of the most important metrics in evaluating the performance of online learning algorithms is the average regret. The average regret is the cumulative loss in reward divided by the horizon length $T$.  This cumulative loss is incurred due to the inaccurate estimation of edge weights and the random nature of the $(\alpha,\gamma)$-approximation oracle invoked. It is worth noting that the loss from a random oracle cannot be reduced even if the true edge weights were known. 

To analyze the performance of our online learning algorithms, we adopt the notion of scaled regret used by \cite{wen2017online}. In particular, let $S^{\text{opt}} := S^{\text{opt}}(w_{\theta^\ast}) =  \arg\max_{S} \{ f(S,w_{\theta^\ast}): |S| \leq K, S\subset \cV\}$ be the optimal size-$K$ seed set with respect to the true parameter $\theta^\ast$, and $S_t$ be the seed set selected at round $t$.  We consider the metric $R^{\alpha \gamma}(T) = \sum_{t=1}^{T} \EE[R_t^{\alpha \gamma}]/T$, where $T$ is the total number of rounds, and $R_t^\eta = f(S^{\text{opt}}, w_{\theta^\ast}) - \frac{1}{\eta} f(S_t, w_{\theta^\ast})$. When $\alpha= \gamma = 1$, $R^{\alpha \gamma}(T)$ reduces to the standard expected average regret $R(T)$.


\subsubsection{Algorithms}
Under the assumptions introduced above, we propose online learning algorithms to learn the true parameter and select seed sets effectively.

Let $\cD^o = \{e_1^o,\ldots,  e_d^o\}$ be the \emph{exploration set}  consisting $d$ diverse edges satisfying Assumption \ref{assump:exploration_nodes}. We partition the time horizon into multiple epochs, each having a number of exploration and exploitation rounds 
depending on a hyper-parameter $q \in \mathbb{Z}_+$. Specifically, the $k$-th epoch consists of $d$ exploration rounds and $k^{q}$ subsequent exploitation rounds. 
More generally, $t_k = 1+ \sum_{m=1}^{k-1}(d+m^{q})$  is the index of the first round of epoch $k$; $\cT_k^o = \{t_k, \ldots,  t_k + d - 1\}$ and $\cT_k^c = \{ t_k + d,\ldots,  t_{k+1} - 1\}$ are the indices of the series of exploration and exploitation rounds in $k$-th epoch, respectively. 

We say that the activation status of a node is \emph{observed} in round $t$ if any of its parent node is activated in the diffusion of round $t$, and \emph{unobserved} otherwise. In round $t$, for any node $v$ whose activation status is observed, denote 
\begin{equation}\label{eq:xbar}\bar{x}_t(v):= \sum_{u\in RP_t(v)} x(u,v),\end{equation}
as the feature for the combined edge weights that are relevant to the activation status of $v$. With $\ind(\cdot)$ denoting the indicator function, let \begin{equation} y_t(v) = \ind\{ \text{$v$ is activated in round $t$}\} = \ind\{ \bar x_t(v) \geq b_v \}.\label{eq:y}\end{equation}

In epoch $k$, our first algorithm runs as follows:
\begin{itemize}
    \item The algorithm first runs $d$ exploration rounds. In the $i$-th exploration round, which has index $t = t_k+i-1$, 
    it selects a single seed node $u^\circ_i := \text{head}(e^\circ_i)$ for $e_i^o \in \cD^\circ$. For each node $v\neq v_i^\circ = \text{tail}(e_i^\circ)$ whose activation status is observed in the current round, we construct 
    $\bar{x}_t(v)$ and $y_t(v)$ as outlined in \eqref{eq:xbar} and \eqref{eq:y}. For node $v_i^\circ = \text{tail}(e_i^\circ)$, we construct \begin{equation}\label{eq:xbar-o} \bar{x}_t(v_i^\circ) = x(e_i^\circ), \;\;\; y_t(v_i^\circ) = \ind\{ \text{$v_i^\circ$ is activated \textit{at step 1} in round $t$}\} = \ind\{ \bar x_t(v_i^\circ) \geq \xi_{v_i^\circ}\}. \end{equation} 
    
    \item After the $d$ exploration rounds in epoch $k$, construct the least square estimate using the observed node activation status as follows: 
    
    
    \[\theta_{k} = \argmin_{\theta}\, \sum_{m = 1}^k \sum_{\tau \in \cT_m^o}\sum_{v\in \cD^o} \| \bar{x}_\tau(v)^\top \theta - y_\tau(v) \|^2 +  \| \theta\|^2.\]
    
    \hspace{0.7cm} Let the covariance matrix and corresponding reward be
    \begin{equation}\label{eq:M_k}
    \M_k :=  \I + \sum_{m = 1}^k \sum_{\tau \in \cT_m^o}\sum_{v\in  \cD^o} \bar{x}_\tau(v)\bar{x}_\tau(v)^\top,\ r_k := \sum_{m = 1}^k \sum_{\tau \in \cT_m^o} \sum_{v \in \cD^o} y_\tau(v) \cdot \bar{x}_\tau(v).
    \end{equation}

    \hspace{0.7cm} $\theta_{k}$ can be represented as
    \begin{align*}\label{theta_ls_est}
        \theta_{k} = \M_{k}^{-1} r_{k}.
    \end{align*}

    \item The algorithm then runs $k^q$ exploitation rounds. 
    In each exploitation round, it invokes the $(\alpha, \gamma)$-oracle on $\cG$ with parameters $\theta_{k}$ to obtain the seed set $S_t$. 
    
    \end{itemize}
    
    Note that in the algorithm presented above, we use only the node observations from exploration rounds to update our belief for $\theta^*$. We do so to simplify the notation in later analysis. In practice, one can use observations from all rounds, including exploitation rounds, to update the belief on $\theta^*$. The theoretical analysis remains the same, but the notation becomes more convoluted.
The complete algorithm is summarized in Algorithm \ref{alg:IM02}.


It is worth noting that to ensure the full observability of the exploration edge $e_i^o$, we only select a single seed node $v_i^o = \text{head}(e_i^o)$ in each exploration round. This way, no other edges will be involved in the attempt to activate $\text{tail}(e_i^o)$ in the first diffusion step of each exploration round. In practice, instead of selecting only $v_i^o$ to make sure edge $e_i^o$'s realization is observed, we can select $v_i^o$ together with a set of seed nodes that are not connected to $u_i^o = \text{tail}(e_i^o)$.

Note also that the number of exploration rounds is fixed to be $d$ while the number of exploitation rounds in the $k$-th epoch is $k^q$. Thus, the ratio between the number of exploration and exploitation rounds decreases as the number of epochs increases. Intuitively, each exploration round incurs regret. As the estimation $\theta_{k}$ gets closer to the true parameter $\theta^*$, we can gradually decrease the number of exploration rounds to reduce the contribution of exploration to the total regret.  At the same time, insufficient exploration could make  $\theta_{k}$ inaccurate, which might lead to sub-optimal seed selection and increased total regret. Thus, a balance of exploration and exploitation is required to achieve desirable total regret. 

In the rest of this section, we provide theoretical analysis on Algorithm \ref{alg:IM02} and derive an expected average scaled regret for it. We also show that asymptotically, the expected average scaled regret is  $\cO(T^{-q/(q+1)})$ where $T$ is the total number of rounds.

\subsection{Regret Analysis}

\begin{theorem} \label{thm:IM01}
Suppose Assumptions  \ref{assump:linear_param} \ref{assump:upepr_bound_sum_features}, \ref{assump:exploration_nodes},
\ref{assump:alpha_beta_oracle}, and
\ref{assump:invariant_Approx_seed_set} hold.
Let $c_k = \sqrt{d \log  (1+ kd )  + d\log(k^q )}   + D $, $\epsilon >0$ be the stability parameter in Assumption \ref{assump:invariant_Approx_seed_set}, and let $k_0 := \min \{k : \frac{c_{k}^2} {\lambda_{\min}^\circ k} \leq \epsilon^2  \}$ where $\lambda_{\min}^o$ is the minimum eigenvalue defined in Assumption \ref{assump:exploration_nodes}. Let $k_t$ be the epoch that round $t$ belongs to, and $t_k$ be the first round of epoch $k$. Let $f^* = f^{opt}(w_{\theta^*})$.
Then the expected average scaled regret of Algorithm \ref{alg:IM02} is 



\begin{equation*}
\frac{1}{T}\sum_{t=1}^T \EE[R_t^{\alpha \gamma}] \leq 
\begin{cases}
  f^* & \text{ if } T < t_{k_0},\\
\frac{ (t_{k_0} -1)  + (k_T-k_0+1)(d+1)}{T} f^*& \text{ if }   T \geq t_{k_0}. 
    \end{cases}
\end{equation*}

\end{theorem}

\proof{Proof sketch.} We give a proof sketch and defer the full proof to Appendix Section \ref{apd:online-lemmas-IM01}. 

Let $c_k = \sqrt{d \log  (1+ kd )  + d\log(k^q )}   + D $, we define the favorable event $\xi_{k}$ as 
\begin{equation}
    \xi_{k}:= \left\{(\theta_m - \theta^*)^T {\M_m} (\theta_m - \theta^*) \leq c_{m}^2,\, \forall\, m=1, 2, \dots k \right\}. \label{good_event}
\end{equation}

The $d$ exploration rounds in each epoch ensure that the minimum eigenvalue of the covariance matrix $\M_k$ grows at least linearly in $k$, where $k$ is the number of epochs. Let $k_T$ be the total number of epochs in a finite horizon with $T$ rounds. The minimum eigenvalue result of $\M_k$ allows us to bound the norm of the difference between $\theta_k$ and the true $\theta^*$ under the favorable event $\xi_{k}$ by $c_{k}/\sqrt{(\lambda_{\min}^\circ k)}$, which is in the order of $\cO(\sqrt{\log(k)/k})$ if we treat $d$ and $q$ as constants. As a result, there is an epoch $k_0$ such that in all epochs $k$ after $k_0$, the distance between our estimated $\theta_k$ and the true parameter $\theta^*$ is smaller than the constant $\epsilon$ in the stability Assumption \ref{assump:invariant_Approx_seed_set}. Thus, under $\xi_{k_T}$, all exploitation rounds after epoch $k_0$ incur 0 regret. The total regret consists of 1) the regret incurred before the end of epoch $k_0$, 2) the regret in the exploration rounds after epoch $k_0$, and 3) the regret in the exploitation rounds after epoch $k_0$ under the complement of the favorable event $\xi_{k_T}$. When $T < t_{k_0}$, the average per round regret is clearly bounded above by $f^*$. When $T > t_{k_0}$, we first show that the favorable event happens with high probability. More specifically, $\PP(\xi_k) \geq 1-1/k^q$. We then bound the expected regret by analyzing 2) and 3) separately. 
\QED

In Theorem~\ref{thm:IM01} above, we establish a finite-time regret analysis for Algorithm \ref{alg:IM02}. Next, we investigate the asymptotic behavior of the regret.

\begin{proposition}\label{prop:asymptotic_regret}
Assume Assumptions  \ref{assump:linear_param} \ref{assump:upepr_bound_sum_features}, \ref{assump:exploration_nodes},
\ref{assump:alpha_beta_oracle}, and
\ref{assump:invariant_Approx_seed_set} hold.
Let $c_k = \sqrt{d \log  (1+ kd )  + d\log(k^q )}   + D $, $\epsilon >0$ be the stability parameter in Assumption \ref{assump:invariant_Approx_seed_set}, and let $k_0 := \min \{k : \frac{c_{k}^2} {\lambda_{\min}^\circ k} \leq \epsilon^2  \}$ where $\lambda_{\min}^o$ is the minimum eigenvalue defined in Assumption \ref{assump:exploration_nodes}. 
Suppose Algorithm \ref{alg:IM02} runs for $T$ rounds. When $T \to \infty$, we have 
$$
 \lim_{T\to \infty} \frac{1}{T}\sum_{t=1}^T \EE[R_t^{\alpha \gamma}] \leq \cO\left(T^{-q/(q+1)}\right).
$$
\end{proposition}
\textit{Proof:} Recall that  $t_k = \sum_{s=1}^{k-1} (d + s^q ) +1 = \Theta( d(k-1) + k^{q+1})$. Let $k_T$ be the index of epoch that round $T$ belongs to. We have $k_T = \Theta(T^{\frac{1}{q+1}})$. Meanwhile,  we have 
$t_{k_0} = \Theta( k_0^{q+1})$.
Therefore, when $T\to \infty$, we have 
\begin{equation*}
\begin{split}
 \lim_{T\to \infty} \frac{1}{T}\sum_{t=1}^T \EE[R_t^{\alpha \gamma}]&= \lim_{T \rightarrow \infty} \frac{\Big ( (t_{k_0} -1)  + (k_T-k_0)(d+1) \Big ) f^* }{T}\\ 
    & \leq \cO\left(\frac{t_{k_0} + T^{\frac{1}{q+1}}}{T}\right) \leq \cO \Big (\frac{k_0^{q+1}}{T} \Big )+\cO\left(T^{-q/(q+1)}\right). 
\end{split}
\end{equation*}
As $k_0 := \min \{k : \frac{c_{k}^2} {\lambda_{\min}^\circ k} \leq \epsilon^2  \}$ is a constant independent of $T$, we have that $\cO (k_0^{q+1}/T) = 0$, which
concludes the proof.

\begin{algorithm}[t]
 \caption{Influence Maximization Linear Node-level Feedback} \label{alg:IM02}
\begin{algorithmic}
\STATE{\bfseries Input:}{ graph $\cG$, seed set cardinality $K$, exploitation parameter $q$, ORACLE, feature vector $x(e)$'s, exploration edges $\cD^o$}
\STATE{\bfseries Initialization:}{ reward $r_0 \leftarrow 0 \in \R^d,$  covariance matrix $\M_0 \leftarrow  \I \in \R^{d\times d}$}
\FOR{epochs $k =1,2,\ldots $} 
\STATE set $t_k = 1+ \sum_{j=1}^{k-1}(d+j^q)$
\STATE set $\M_k \leftarrow \M_{k-1}$, $r_k \leftarrow r_{k-1}$
\FOR{$t=t_k,\ldots, t_k + d - 1$}
    \STATE set $i = t-t_k+1$
    \STATE choose $S_t = \{\text{head}(e_i^o)\}$ where $e_i^o \in \cD^o$
    \FOR{$v \in \cV$ such that observation status of $v$ is observed}
    \STATE construct $\bar x_t(v), y_t(v)$ as in Eqs.\eqref{eq:xbar}, \eqref{eq:y}, \eqref{eq:xbar-o}
    \STATE update $\M_k \leftarrow \M_k + \bar x_t(v) \bar x_t^T(v) $ and $r_k \leftarrow r_k + \bar x_t(v) y_t(v)$
    \ENDFOR 
\ENDFOR 
\STATE set $\theta_{k}= \M_{k}^{-1} r_{k}$
\FOR{$t=t_k + d,\ldots, t_{k+1} - 1$}
\STATE choose $S_t \in ORACLE(\cG,K,w_{\theta_{k}})$
\ENDFOR
\ENDFOR 
 \end{algorithmic}
\end{algorithm}

\section{Learning in Negativity-Aware LT-N Model} \label{sec:ON-LTN}

For our LT-N model, we not only need to learn $\theta^* \in \mathbb{R}^d$ associated with the edge weights but also $\beta^* \in \mathbb{R}^{d'}$ associated with the autonomy factors $q^+(v)$ and $q^-(v)$ for each node $v \in \cV$. For learning $\theta^*$, we use the same observation as learning the edge weights in the classical LT-model, as discussed in Section \ref{sec:ON}. Namely, we use the observed unsigned activation of nodes in each time step to update our belief on $\theta^*$. We can thus use the results that we developed in the previous section. More specifically, we have that in any epoch $k$, if $\theta^* \in \cC_k$, then $\|\theta^* - \theta_k \|\leq \cO(\sqrt{\log(k)/k})$.

As for learning $\beta^*$, we need to use the observations on the signs of the activation. Recall that for each $v \in \cV$, we have two known feature vectors $x_+(v), x_-(v) \in \mathbb{R}^{d'}$ such that the autonomy factors $q^+(v) = x^{\top}_+(v)\beta^*, q^-(v) = x^{\top}_-(v)\beta^*$. Also, $r(v) = q^+(v) + q^-(v)$. Furthermore,
given that $v$ becomes active in $\tau$, the probability of it being positive is 
\begin{align*}
& q^+(v) + [1-r(v)] \frac{\sum_{u \in \iN(v) \cap (B^+_{\tau-1}\setminus B_{\tau-2})} w(u, v)}{\sum_{u \in \iN(v) \cap (B_{\tau-1}\setminus B_{\tau-2})} w(u, v)} \\
& =  x^{\top}_+(v)\beta^* + (1- x^{\top}_+(v)\beta^* - x^{\top}_-(v)\beta^*)\cdot \frac{\sum_{u \in \iN(v) \cap (B^+_{\tau-1}\setminus B_{\tau-2})} x(u,v)^\top \theta^*}{\sum_{u \in \iN(v) \cap (B_{\tau-1}\setminus B_{\tau-2})}x(u,v)^\top \theta^*}.
\end{align*}


Recall that for each node $v \in \cV$, we use $\tau_t(v)$ to denote the time step at which node $v$ becomes activated in round $t$. When $v\in S_t$, we have that $\tau_t(v)=0$. If $v$ is not activated in round $t$, then we set $\tau_t(v) = \infty$. 

For simplicity of notation, we use $p_t^+(v,\theta,\beta)$ to denote the random probability of node $v$ becoming positive in round $t$ given that it is activated in round $t$, the edge weights are linear with respect to $\theta$, and the autonomy factors are linear  with respect to $\beta$. Specifically,
\begin{align*}
p_t^+(v,\theta,\beta) = x^{\top}_+(v)\beta + \Big (1- x^{\top}_+(v)\beta - x^{\top}_-(v)\beta \Big )\cdot \frac{\sum_{u \in \iN(v) \cap (B^+_{\tau_t(v)-1}\setminus B_{\tau_t(v)-2})} x(u,v)^\top \theta}{\sum_{u \in \iN(v) \cap (B_{\tau_t(v)-1}\setminus B_{\tau_t(v)-2})}x(u,v)^\top \theta}.
\end{align*}
The randomness of $p_t^+(v,\theta,\beta)$ comes from the randomness of $B_{\tau_t(v)}$'s and $B_{\tau_t(v)}^+$'s in round $t$, where $\tau_t(v)$ is the random time step in round $t$ when $v$ is activated.

In this section, we first state and justify several additional assumptions that extend the ones in Section \ref{sec:ON} for learning with the classical LT model. We then detail our learning algorithm for LT-N and prove a regret bound for it.

\subsection{Technical Assumptions}
\begin{assumption}[Linear parameterization]\label{assump:linear_param_2}
    There exists $\beta^* \in \mathbb{R}^{d'}$, $\|\beta^*\| \leq D'$ such that the true autonomy factors are $q^+(v) = x^{\top}_+(v)\beta^*, q^-(v) = x^{\top}_-(v)\beta^*$ with $q^+(v), q^-(v)\geq 0$, and $r(v) = q^+(v) + q^-(v) \leq 1$ for all $v \in \cV$.
\end{assumption}
With a separate vector $\beta^*$, we are assuming that the autonomy factors are linear functions of individual user characteristics. As before, we use $\beta$ to denote a generic vector in $\R^{d'}$ and refer to it as the \emph{parameter}. We refer to $\beta^\ast$ as the \emph{true parameter}. We denote the \emph{estimated parameter} after the exploration rounds of epoch $k$ as $\beta_k$.

One of our key ideas for learning with LT is to make sure that the features of observed edges are diverse enough to allow enough information to be collected on all directions of $\theta_t$,  so that $\theta_t \rightarrow \theta^\ast $ as $t \rightarrow \infty$. For LT-N, we want to make sure that the autonomy features of activated nodes are diverse enough to allow information to be collected on all directions of $\beta_t$, so that $\beta_t \rightarrow \beta^\ast$ as $t \rightarrow \infty$. We thus make the following assumption:
\begin{assumption}[Feature diversity] \label{assump:exploration_nodes-2}
        There exists $d'$ nodes $v_i^{auto}, 1\leq i \leq d'$, such that each $v_i^{auto}$ has at least one parent, and matrix $\sum_{i=1}^{d'}x_{
    -}(v_i^{auto})x_{-}(v_i^{auto})^\top$ is positive definite. In other words, the minimum eigenvalue $\lambda_{\min}^{auto}: = \lambda_{\min} \Big (\sum_{i=1}^{d'} x_{-}(v_i^{auto})x_{-}(v_i^{auto})^\top \Big )$ is strictly greater than 0. 
\end{assumption}
This assumption should be easy to satisfy as the dimension of the features is usually much smaller than the total number of nodes, that is, $d' \ll |\cV|$.

Let $K$ be the maximum number of seed nodes that can be chosen in each round. Let $f^+(S,w,q^\pm)$ be the expected total positive influence given seed set $S$, edge weights $w$, and autonomy factors $q^\pm = (q^+, q^-)$. Denote $S^{\text{opt}}(w,q^\pm) = \arg\max \{ f^+(S,w,q^\pm): |S|\leq K,\ S\subseteq \cV \}$ and $f^{\text{opt}}(w,q^\pm) = \max \{ f^{+}(S,w,q^\pm): |S|\leq K,\ S\subseteq \cV \}$. 
From Theorem \ref{thm:sig^+_mono_subm}, we know that $f^+(\cdot, w, q^\pm)$ is monotone submodular. Therefore, a greedy algorithm with exact evaluation of $f^+(\cdot, w, q^\pm)$ returns a $(1-1/e)$-approximation solution. Since evaluating $f^+(\cdot, w, q^\pm)$ is $\#$-P hard, we assume access to an approximation oracle: 
\begin{assumption}[Approximation oracle]\label{assump:alpha_beta_oracle-2} Let $\epsilon >0$ and $\alpha = 1-1/e - \epsilon$.
There exists an efficient, possibly random $(\alpha,\gamma)$-\emph{oracle} that takes $\cG$, $w$, $q^\pm$, $K$, and outputs a seed set $\tilde S$ such that $f^+(\tilde S,w,q^\pm)\geq \alpha \cdot f^{\text{opt}}(w,q^\pm)$ with probability at least $\gamma$.
\end{assumption}
As in the case of classical LT, the reverse reachable set method in \cite{tang2014influence} can be easily extended to obtain such an oracle. 

We also impose assumptions on the solution stability of the network. Consider the solution returned by the approximation oracle around $\theta^\ast$ and $\beta^\ast$.
Use $q_{\beta}$ to represent the autonomy factors $q_{\beta}^+(v) = x_+(v)^\top \beta$ and $q_{\beta}^-(v) = x_-(v)^\top \beta$. Define the \textit{$\alpha$-approximation seed sets} with respect to $\theta \in \mathbb{R}^d$ and $\beta \in \mathbb{R}^{d'}$ as $\cA(\theta, \beta, \alpha) = \{ S\subseteq\cV: f^+(S,w_\theta, q_\beta) \geq  \alpha\cdot f^{\text{opt}}(w_\theta, q_\beta)\}$. That is, $\cA(\theta,\beta,\alpha)$ is the set of seed sets whose expected positive influence with respect to influence probabilities $w_\theta(e) = x(e)^\top \theta$ and autonomy factors $q_{\beta}^+(v) = x_+(v)^\top \beta$ and $q_{\beta}^-(v) = x_-(v)^\top \beta$ is at least $\alpha$ times that of an optimal seed set. Our next assumption states that the set of  $\alpha$-approximation sets and the optimal seed set are invariant under small perturbations.
\begin{assumption}[Stability:LT-N]\label{assump:invariant_Approx_seed_set_LTN} 
    There exist constants $\epsilon_\theta,\epsilon_\beta>0$ such that for any $\theta \in \mathbb{R}^d$ and $\beta \in \mathbb{R}^{d'}$ that satisfies $\| \theta-\theta^*\|\leq \epsilon_\theta$ and $\|\beta - \beta^* \|\leq \epsilon_\beta$, we have $\cA(\theta,\beta,\alpha) = \cA(\theta^*, \beta^*,\alpha)$. Moreover, there exists a seed set $S^{\cA}$ such that $S^{\cA} \in \argmin\{f^{+}(S,w_{\theta},q_{\beta}): S\in \cA(\theta,\beta,\alpha)\}$ for all $(\theta,\beta)$ such that $\| \theta-\theta^*\|\leq \epsilon_\theta$ and 
    $\|\beta - \beta^* \|\leq \epsilon_\beta$. 
\end{assumption}
Using similar argument as in the proof of Theorem \ref{thm:2}, it is easy to see that Assumption \ref{assump:invariant_Approx_seed_set_LTN} holds with probability 1 if $\alpha$ is uniformly sampled from an interval $[1-1/e-\epsilon_1, 1-1/e - \epsilon_2]$ with $\epsilon_1 > \epsilon_2$.

\subsection{Algorithms}
Under Assumptions \ref{assump:linear_param},
 \ref{assump:upepr_bound_sum_features},
\ref{assump:exploration_nodes},
\ref{assump:linear_param_2},
\ref{assump:exploration_nodes-2},
\ref{assump:alpha_beta_oracle-2}, and
\ref{assump:invariant_Approx_seed_set_LTN} introduced above, we propose online learning algorithms to learn the true parameter and select seed sets effectively.

Let $\cD^o = \{e_1^o,\ldots,  e_d^o\}$ be the \emph{edge exploration set}  consisting $d$ diverse edges satisfying Assumption \ref{assump:exploration_nodes}.  Let $\cD^{auto} = \{v_1^{auto}, \ldots, v_{d'}^{auto}\}$ be the \emph{node exploration set} consisting $d'$ nodes that satisfy the autonomy feature diversity part of Assumption \ref{assump:exploration_nodes-2}. 

Let $q \in \mathbb{Z}^+$ be a hyper-parameter of the algorithm. We partition the time horizon into multiple  epochs, each having several exploration and exploitation rounds. Specifically, the $k$-th epoch consists of $d+d'$ exploration and $k^q$ subsequent exploitation rounds. 
More generally, $t_k = 1+ \sum_{m=1}^{k-1}(d+d'+k^q)$ is the index of the first round of epoch $k$; $\cT_k^o = \{t_k, \ldots,  t_k + d +d'- 1\}$ and $\cT_k^c = \{ t_k + d +d',\ldots,  t_k + d + d' + k^{q} -1\}$ are the series of exploration and exploitation rounds in the $k$-th epoch, respectively. We further define $\cT_k^{od} = \{t_k, \ldots,  t_k + d - 1\}$ and $\cT_k^{od'} = \{t_k + d, \ldots,  t_k + d +d' - 1\}$.

We say that the activation status of a node is \emph{observed} in round $t$ if any of its parent node is activated in the diffusion of round $t$, and \emph{unobserved} otherwise. In round $t$ for any node $v$ whose activation status is observed, denote 
\begin{equation}\bar{x}_t(v):= \sum_{u\in RP_t(v)} x(u,v), \label{eq:xbar-2}\end{equation}
as the feature for the combined edge weights that are relevant to the activation status of $v$, and let \begin{equation}y_t(v) = \ind\{ \text{$v$ is activated in round $t$}\} = \ind\{ \bar x_t(v) \geq b_v \}.\label{eq:y-2}\end{equation}
Use $\cV_t^o$ to denote the set of observed nodes in round $t$. Use $\cA_t^o$ to denote the set of activated nodes in round $t$.

In epoch $k$, our algorithm with hyper-parameter $q$ runs as follows:
\begin{itemize}
    \item The algorithm first runs $d$ exploration rounds. In the $i$-th exploration round which has round index $t = t_k+i-1$, 
    it selects a single seed node $u^\circ_i := \text{head}(e^\circ_i)$ for $e_i^o \in \cD^\circ$. For each node $v\neq v_i^\circ = \text{tail}(e_i^\circ)$ whose activation status is observed in the current round, we construct $\bar{x}_t(v)$ and $y_t(v)$ as outlined in \eqref{eq:xbar-2} and \eqref{eq:y-2}. For node $v_i^\circ = \text{tail}(e_i^\circ)$, we construct 
    \begin{equation}\label{eq:xbar-o-2}\bar{x}_t(v_i^\circ) = x(e_i^\circ), \;\;\; y_t(v_i^\circ) = \ind\{ \text{$v_i^\circ$ is activated \textit{at step 1} in round $t$}\} = \ind\{ \bar x_t(v_i^\circ) \geq \xi_{v_i^\circ}\}.\end{equation} 
    
    \item After the $d$ exploration rounds in epoch $k$, construct the least squares estimate using the observed node activation status as follows:
    
    \[ \theta_{k} = \argmin_{\theta}\, \sum_{m = 1}^k \sum_{\tau \in \cT_m^{od}}\sum_{v\in \cD^o} \| \bar{x}_\tau(v)^\top \theta - y_\tau(v) \|^2 +  \| \theta\|^2.\]
    
    \hspace{0.7cm} Let the covariance matrix and corresponding reward be
    \begin{equation*}
    \M_k :=  \I + \sum_{m = 1}^k \sum_{\tau \in \cT_m^{od}}\sum_{v\in \cD^o}  \bar{x}_\tau(v)\bar{x}_\tau(v)^\top,\ r_k := \sum_{m = 1}^k \sum_{\tau \in \cT_m^{od}}\sum_{v\in \cD^o}  y_\tau(v) \cdot \bar{x}_\tau(v).
    \end{equation*}

    \hspace{0.7cm} $\theta_{k}$ can be represented as
    \begin{align*}
\theta_{k} = \M_{k}^{-1} r_{k}.
    \end{align*}
    \item The algorithm then runs $d'$ more exploration rounds for autonomy factors. In the $i$-th such exploration round, which has round index $t = t_k+d+i-1$, 
    it selects $\min\{|\cN^{in}(v_i^{auto})|,K\}$ parent nodes of node $v_i^{auto} \in \cD^{auto}$ (that have the smallest indices) as the seed set. More specifically, we fix a parent set of $v_i^{auto}$ 
    with cardinality no more than $K$, and 
    denote it by $N(v_i^{auto})$. We also define $$w_i^{auto} := \sum_{u \in N(v_i^{auto})}w(u,v_i^{auto}). $$ Clearly, if $N(v_i^{auto})$ is chosen as the seed set, the probability that $v_i^{auto}$ is activated at time step 1 equals $w_i^{auto}$.  
     Correspondingly, if node $v_i^{auto}$ is activated in step 1 of the diffusion process, then the probability of it turning positive is $$x^{\top}_+(v_i^{auto})\beta^* + (1- x^{\top}_+(v_i^{auto})\beta^* - x^{\top}_-(v_i^{auto})\beta^*) \cdot 1 = 1- x^{\top}_-(v_i^{auto})\beta^*.$$ This is because all active parent nodes of node $v_i^{auto}$ before step 1 are positively activated as seed nodes. Observe the signs of the activated nodes in the round. For each $v$ that has been activated in round $t$, let \begin{equation*}y^+_t(v) = \ind\{ \text{$v$ is positively activated in round $t$}\}.\end{equation*}
    
    \item After the additional $d'$ exploration rounds, obtain the least squares estimate for $\beta^*$ using the observed signs of activated nodes in the $d'k$ exploration rounds of autonomy factors in the first $k$ epochs 

    More specifically, let the covariance matrix and the corresponding reward be
    \begin{equation*}
    \begin{split}
    \V_k & :=  \I + \sum_{m = 1}^k\sum_{t \in \cT_m^{od'}}\sum_{v\in \cD^{auto}} \ind \{ v \text{ is activated in step 1 of round $t$}\} \cdot x_{-} (v ) x_{-}^\top (v )
  ,   \\
    \ s_k & := \sum_{m = 1}^k\sum_{t \in \cT_m^{od'}}\sum_{v\in \cD^{auto}}\ind \{ v \text{ is activated in step 1 of round $t$}\} \cdot x_{-} (v )  (1-y_t^{+}(v)).
    \end{split}
    \end{equation*}
    $\beta_k$ can be represented as
    $$\beta_k = \V_k^{-1}s_k.$$
    \item The algorithm then runs $k^q$ exploitation rounds. In each exploitation round $t$, 
    it invokes the $(\alpha, \gamma)$-oracle on $\cG$ with inputs $w_{\theta_{k}}$ and $q_{\beta_{k}}$ to obtain the seed set $S_t$. 

\end{itemize}

We summarize the steps explained above in Algorithm \ref{alg:IM03}. 
\begin{algorithm}[t]
 \caption{Influence Maximization LT-N} \label{alg:IM03}
\begin{algorithmic}
\STATE{\bfseries Input:}{ graph $\cG$, seed set cardinality $K$, exploitation parameter $q$, ORACLE, feature vector $x(e)$'s, $x_+(v), x_-(v)$'s, $\cD^o$, }.
\STATE{\bfseries Initialization:}{ reward $r_0 \leftarrow 0 \in \R^d,$  corvariance matrix $\M_0 \leftarrow  \I \in \R^{d\times d}$}
\FOR{epochs $k =1,2,\ldots $} 
\STATE set $t_k = 1+ \sum_{j=1}^{k-1}(d+d'+j^q)$
\STATE set $\M_k \leftarrow \M_{k-1}$, $r_k \leftarrow r_{k-1}$, $\V_k \leftarrow \V_{k-1}$, $s_k \leftarrow s_{k-1}$.
\FOR{$t=t_k,\ldots, t_k + d - 1$}
    \STATE set $i = t-t_k+1$
    \STATE choose $S_t = \{\text{head}(e_i^o)\}$ where $e_i^o \in \cD^o$.
    \STATE update $\M_k \leftarrow \M_{k} + \bar x_t(v) \bar x_t^T(v) $ and $r_k \leftarrow r_k + \bar x_t(v) y_t(v)$.
\ENDFOR 
\STATE set $\theta_{k}= \M_{k}^{-1} r_{k}$
\FOR{$t=t_k + d,\ldots, t_k + d +d' - 1$}
    \STATE set $i = t-t_k-d+1$
    \STATE choose $S_t = N(v_i^{auto})$
    \IF{$v_i^{auto}$ is activated in step 1 of the diffusion process}
        \STATE update 
$
\V_k \leftarrow \V_{k} + x_{-}(v)  x_{-}^\top (v)
$ and $s_k \leftarrow s_{k} + x_{-}(v) (1-y_t^{+}(v))$
\ENDIF
\ENDFOR 
\STATE set $\beta_{k} = \V_k^{-1}s_k$.

\FOR{$t=t_k + d + d',\ldots, t_{k+1} - 1$}
\STATE 
choose $S_t \in ORACLE(\cG,K,w_{\theta_{k}},q_{\beta_{k}})$
\ENDFOR
\ENDFOR 
 \end{algorithmic}
\end{algorithm}
\subsection{Regret Analysis}
In Section \ref{sec:ON}, we show that Algorithm \ref{alg:IM02} achieves an asymptotic expected average scaled regret of $\cO(T^{-q/(q+1)})$ for classical LT models. However, when autonomy factors are involved, the performance of Algorithm \ref{alg:IM03} for LT-N models remains unexplored. 
In this part, we present the analysis of the performance of Algorithm \ref{alg:IM03}. We adopt similar notation as those in Section \ref{sec:ON}. We define $R_t^{\alpha \gamma}$ as the $\alpha \gamma$-scaled regret incurred in round $t$, and define $R^{\alpha \gamma}(T)$ as the cumulative regret over $T$ rounds. That is, $R^{\alpha \gamma}(T) = \sum_{t=1}^T R_t^{\alpha \gamma}$. We present an upper bound on the average regret $R^{\alpha \gamma}(T)/T$ in Theorem \ref{thm:IM03} below, the proof of which can be found in Appendix Section \ref{apd:online-lemmas-IM03}.

\begin{theorem}\label{thm:IM03}
Suppose Assumptions \ref{assump:linear_param},
 \ref{assump:upepr_bound_sum_features},
\ref{assump:exploration_nodes},
\ref{assump:linear_param_2},
\ref{assump:exploration_nodes-2},
\ref{assump:alpha_beta_oracle-2}, and
\ref{assump:invariant_Approx_seed_set_LTN} hold. Let $c_k = \sqrt{d \log  (1+ kd )  + d\log(k^q )}   + D $, $\kappa_k = \sqrt{d' \log \left(1+ kd' \right) + d' \log (k^q) }  + D_{\beta}$, let $\epsilon_\theta,\epsilon_{\beta} >0$ be the stability parameters in Assumption \ref{assump:invariant_Approx_seed_set_LTN}, let $\varepsilon_k = \sqrt{\frac{q \log k }{2k}}$, and let $k_0 := \min \{k : \frac{c_{k}^2} {\lambda_{\min}^\circ k} \leq \epsilon_{\theta}^2, \frac{\kappa_k^2}{(k\cdot  w_{\min}^{auto}     - \varepsilon_k) \lambda_{\min}^{auto}}  \leq \epsilon_{\beta}^2\}$ where $\lambda_{\min}^o$ and $\lambda_{\min}^{auto}$ are the minimum eigenvalues defined in Assumptions \ref{assump:exploration_nodes} and \ref{assump:exploration_nodes-2}, respectively. Let $k_t$ be the epoch that round $t$ belongs to, and $t_k$ be the first round of epoch $k$. Let $f^* = f^{\text{opt}}(w_{\theta^*},q_{\beta^*})$.
Then the expected average scaled regret of Algorithm \ref{alg:IM03} over $T$ rounds is  
\begin{equation*}
\frac{1}{T}\sum_{t=1}^T \EE[R_t^{\alpha \gamma}] \leq 
\begin{cases}
  f^* & \text{ if } T < t_{k_0},\\
\frac{(t_{k_0} -1)  + (k_T-k_0)(d+2d'+1)}{T} f^*& \text{ if }   T \geq t_{k_0}, 
    \end{cases}
\end{equation*}
\end{theorem}

\begin{proposition}
Assume Assumptions \ref{assump:linear_param},
 \ref{assump:upepr_bound_sum_features},
\ref{assump:exploration_nodes},
\ref{assump:linear_param_2},
\ref{assump:exploration_nodes-2},
\ref{assump:alpha_beta_oracle-2}, and
\ref{assump:invariant_Approx_seed_set_LTN} hold. 
As the number of rounds $T\to \infty$, Algorithm \ref{alg:IM03} has an asymptotic expected average scaled regret 
$$
 \lim_{T\to \infty} \frac{1}{T}\sum_{t=1}^T \EE[R_t^{\alpha \gamma}] \leq \cO\left(T^{-q/(q+1)}\right).
$$
\end{proposition}
The above result could be derived by similar analysis as in Proposition \ref{prop:asymptotic_regret}. We omit the details here to avoid repetition.

\section{Numerical Experiments} \label{NE}
To numerically test the performance of various seeding strategies, we conduct experiments on a Twitter subnetwork with 232 nodes and 3090 directed edges. The complete directed graph with 232 nodes has 53,592 directed edges. Thus in our subnetwork, around 6 percent of all possible edges are present. We obtain the network structure from the SNAP dataset by \cite{SNAP}. The algorithms we test are summarized in Table \ref{tab:nomen1}. We also explain each algorithm in detail in the section on experimental setup.

\begin{center}
	\begin{table}[ht]
		\caption{Summary of algorithms tested in the numerical experiments}
		\label{tab:nomen1}
		\centering
\begin{tabular}{||l|l||}
	
	 \hline 
	\begin{tabular}{c}
		\textbf{Algorithm}\\
	\end{tabular} &
	\begin{tabular}{l}
		\textbf{Meaning}\\
	\end{tabular}
	\\
	\hline 
	\begin{tabular}{c}
		{\tt{bgg\_dgr}}\\
	\end{tabular} &
	\begin{tabular}{l}
		Seed five nodes that have maximum out-degrees in each round 
	\end{tabular}
	\\
	\hline 
	\begin{tabular}{c}
		{\tt{rdm}}\\
	\end{tabular} &
	\begin{tabular}{l}
		Seed five different randomly sampled nodes in each round\\
	\end{tabular}
	\\
	\hline 
	\begin{tabular}{c}
		{\tt{grd\_kw}}\\
	\end{tabular} &
	\begin{tabular}{l}
		Seed five nodes selected by approximate greedy oracle with known \\true edge weights in each round\\
	\end{tabular}
	\\
	\hline 
	\begin{tabular}{c}
		{\tt{grd\_explr\_q=1}}\\
	\end{tabular} &
	\begin{tabular}{l}
		Algorithm \ref{alg:IM02} with $q=1$\\
	\end{tabular}
	\\
	\hline 
	\begin{tabular}{c}
		{\tt{grd\_explr\_q=2}}\\
	\end{tabular} &
	\begin{tabular}{l}
		Algorithm \ref{alg:IM02} with $q = 2$\\
	\end{tabular}
	\\
	\hline 
	\begin{tabular}{c}
		{\tt{grd\_explr\_q=3}}\\
	\end{tabular} &
	\begin{tabular}{l}
		Algorithm \ref{alg:IM02} with $q = 3$\\
	\end{tabular}
	\\
	\hline 
	\begin{tabular}{c}
		{\tt{grd\_splt}}\\
	\end{tabular} &
	\begin{tabular}{l}
		A comparison learning and seeding algorithm that assigns node observations \\equally to the contributing edges when updating beliefs\\
	\end{tabular}
	\\ \hline 
\end{tabular}

\end{table}

\end{center}

To generate edge feature vectors, we use the node2vec algorithm proposed by \cite{DBLP:journals/corr/GroverL16} to first construct node feature vectors, and then use the element-wise multiplication of the head node's and tail node's vectors of each edge as the corresponding edge feature vector. We then randomly perturb the resulting vectors so that they become more diverse. We set the feature dimension to be 5, and we hand-pick a theta vector that has 3 positive entries and 2 negative entries. The 2-norm of this theta vector is around 1.89. 

We then construct the edge weights using the dot product of the edge feature vectors and our theta vector. Whenever we have a negative weight, we replace it with 0. Our diffusion model assumes that the sum of incoming weights to each node is between 0 and 1. Therefore, for each node, we sum up its incoming weights. If this sum is greater than 1, we scale down the feature vectors and the weights of the incoming edges uniformly by the sum. Also, we scale down the feature vectors and weights of out-going edges from several high-degree nodes, so that the optimal seed set is unlikely to be just the set of highest-degree nodes. In this way, the learning algorithm needs to learn a good estimator of theta to be able to select good seed sets. 

We treat the feature vectors and weights obtained using the process described above as the ground truth. Note that the linear generalization of edge weights might not be perfect, as we have cropped the negative weights in an early step. We pick a set of 5 edges $(u_s, v_s), s = 1, 2, ..., 5$ as our exploration set and check that they satisfy Assumption \ref{assump:exploration_nodes}, the feature diversity assumption, as desired.

We simulate the learning and seeding process with Algorithm \ref{alg:IM02} for $q = 1, 2, 3$ ({\tt{grd\_explr\_q=1, grd\_explr\_q=2, grd\_explr\_q=3}}) for 615 rounds (which corresponds to 30 epochs of running Algorithm \ref{alg:IM02} with $q = 1$). In each epoch $m$, we first do 5 exploration rounds, in which we seed $u_s$ along with 4 highest-degree nodes among the ones not pointing to $v_s$, and observe the 1-step activation status of $v_s$ to update the weight estimates. We then do $m^q$ rounds of exploitation, in each round of which we first compute the weight estimates and then feed the estimates into our approximate-greedy oracle to get a seed set of cardinality 5. We then observe the resulting diffusion and update the weight estimates accordingly. For other baseline algorithms that do not involve exploration rounds, we do 615 rounds of seeding using the respective seeding strategies. 

\begin{figure}
    \centering
    \includegraphics[scale=0.45]{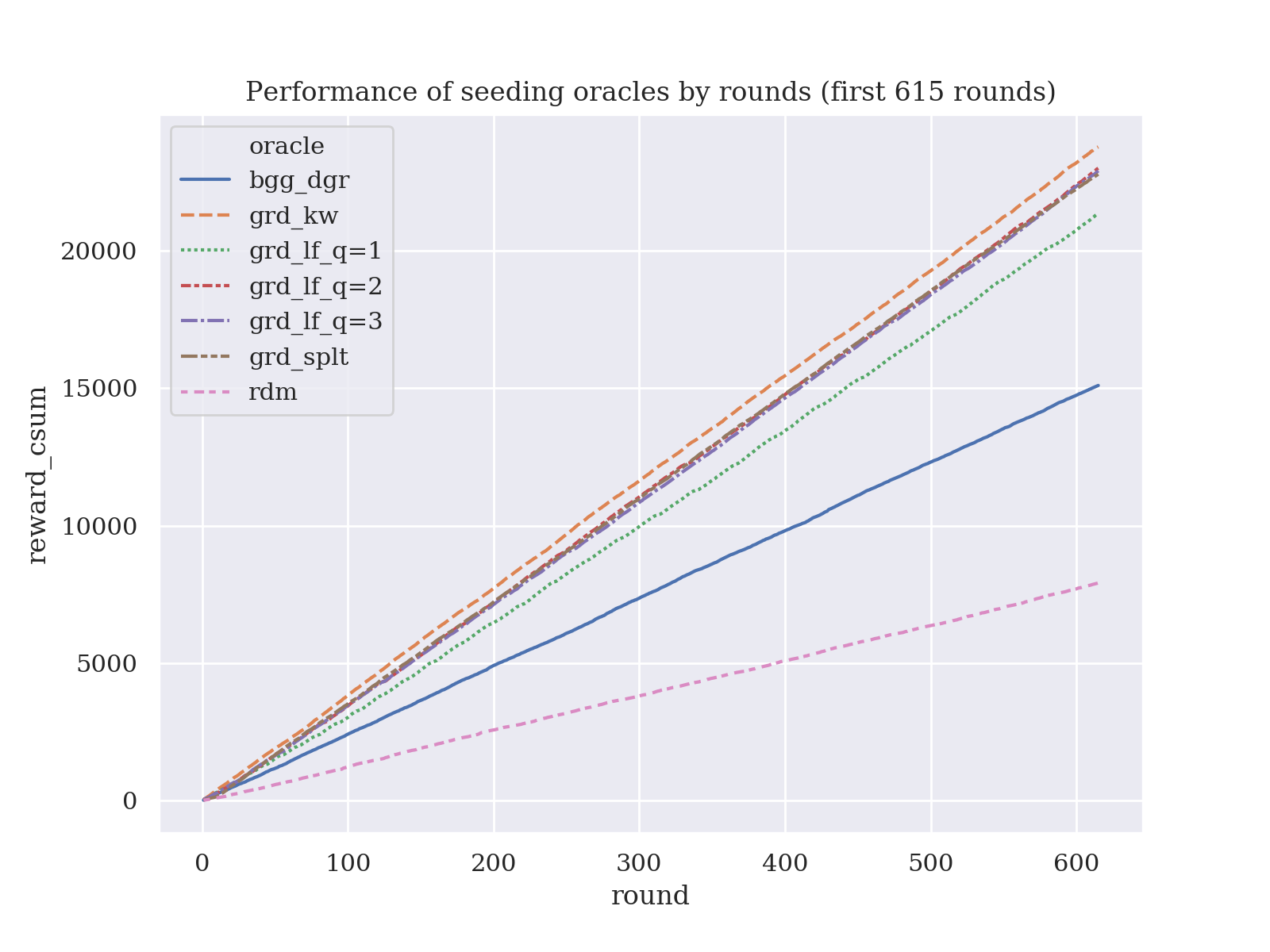}
    \caption{Cumulative reward by round over all 615 rounds.}
    \label{fig:3}
\end{figure}

\begin{figure}
    \centering
    \includegraphics[scale=0.4]{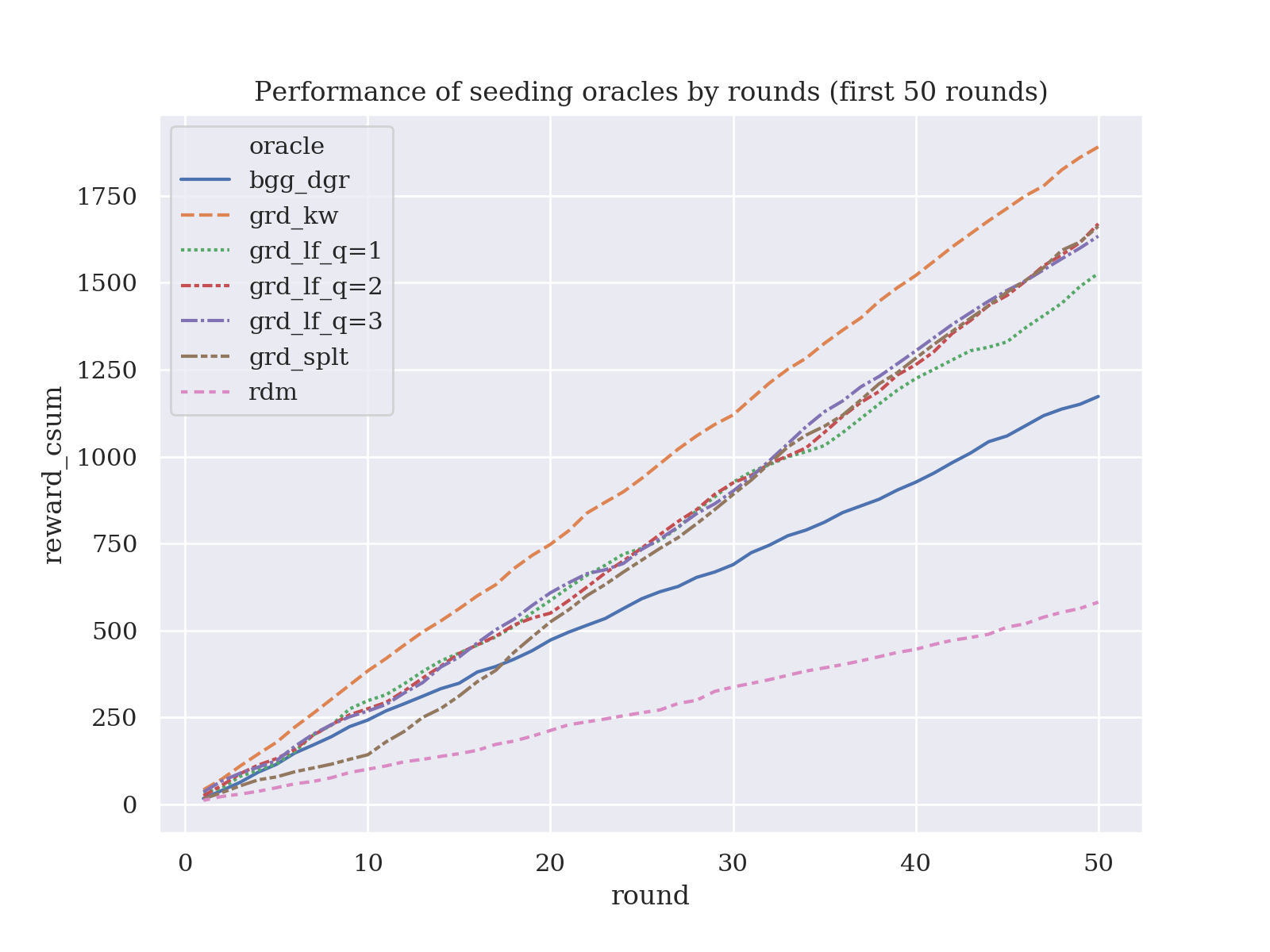}
    \includegraphics[scale=0.4]{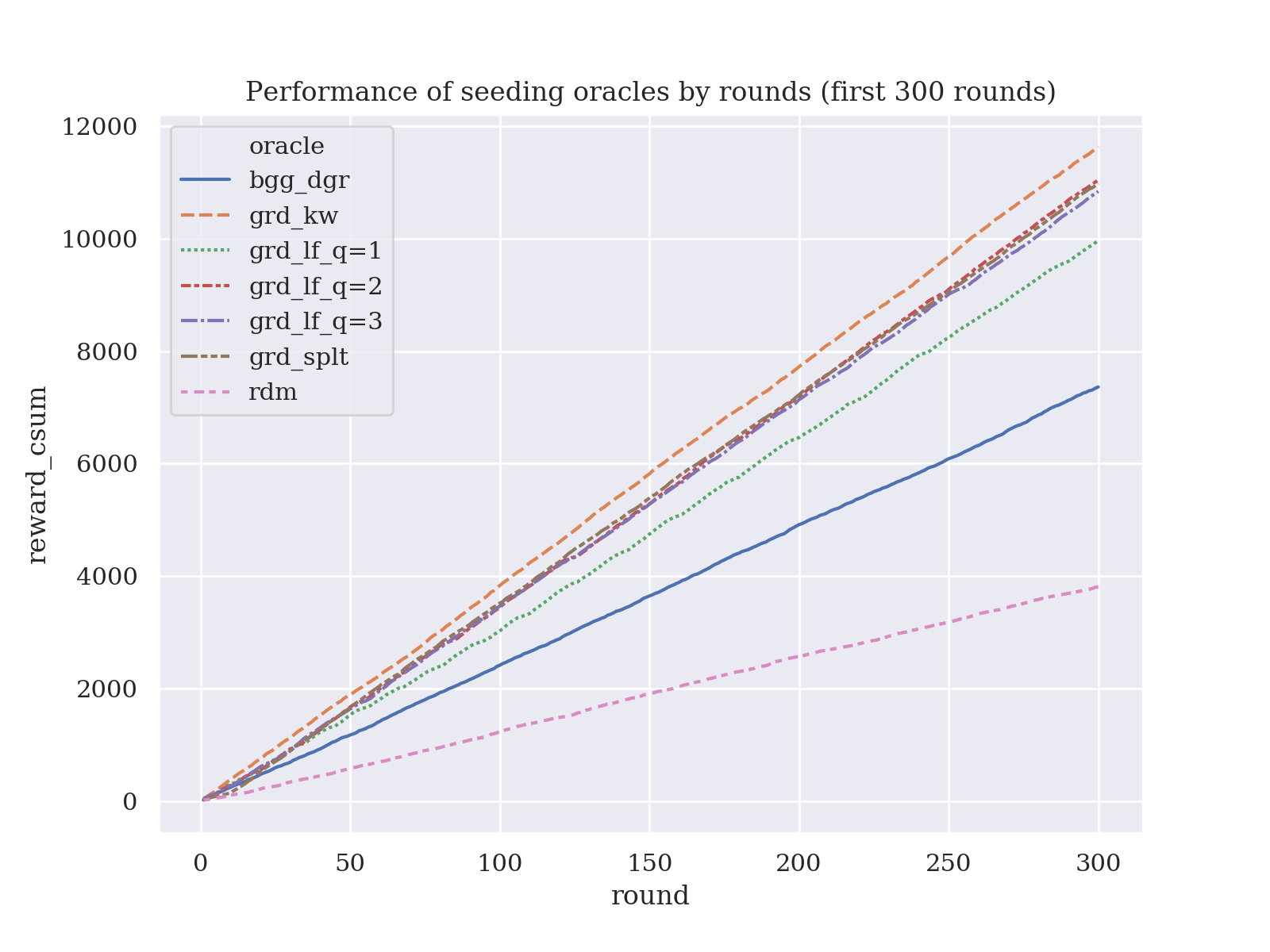}
    \caption{Cumulative reward by round (zoomed in first 50 rounds and first 300 rounds).}
    \label{fig:4}
\end{figure}

We compare the sum of round reward achieved by {\tt{grd\_explr\_q=1}}, {\tt{grd\_explr\_q=2}}, and {\tt{grd\_explr\_q=3}} with {\tt{grd\_kw}}, the approximate greedy oracle that knows the true weights, as well as another learning and seeding algorithm, {\tt{grd\_splt}}. In the latter, when a node $v$ is activated in round $t$, we attribute $\frac{1}{|RP_t(v)|}$ of its activation to each relevant edge $e = (u, v), u \in RP_t(v)$. We then update the theta estimation using individual edge feature vectors instead of the sum of edge feature vectors of the relevant edges. Additionally, we test the oracle that randomly samples a seed set of cardinality 5 in each round ({\tt{rdm}}), and the oracle that samples the set of 5 highest degree nodes in each epoch ({\tt{bgg\_dgr}}) in each round.

\begin{figure}
    \centering
    \includegraphics[scale=0.45]{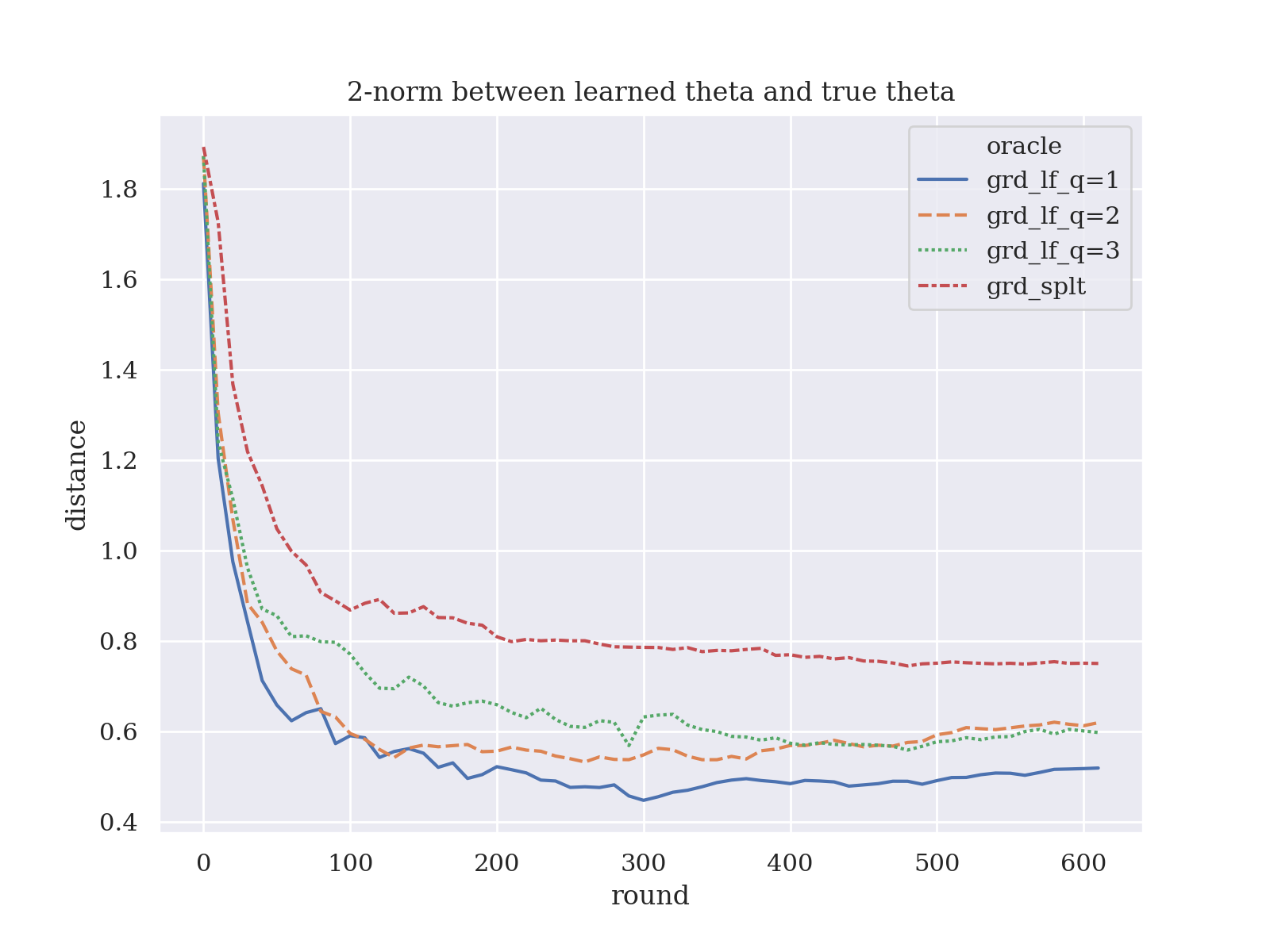}
    \caption{2-norm between true theta and learned theta.}
    \label{fig:5}
\end{figure}

The results are summarized in Figure \ref{fig:3}, \ref{fig:4}, an \ref{fig:5}. Each plot is produced by averaging over 5 independent simulations. As we can see from Figure \ref{fig:3}, {\tt{grd\_explr\_q=2}}, {\tt{grd\_explr\_q=3}}, and {\tt{grd\_splt}} achieve similar performance over the first 615 rounds, and their performance are very close to {\tt{grd\_kw}} which knows the true influence probabilities. {\tt{grd\_explr\_q=1}} has worse performance, but its seeding results are still considerably better than the baselines {\tt{rdm}} and {\tt{bgg\_dgr}}. We also plot the distance between the learned theta and the true theta over all 615 rounds for {\tt{grd\_splt}},  {\tt{grd\_explr\_q=1}}, and {\tt{grd\_explr\_q=2}}, {\tt{grd\_explr\_q=3}} in Figure \ref{fig:5}. We see that with exploration rounds, the theta learned by {\tt{grd\_explr\_q=1}, {\tt{grd\_explr\_q=2}}, and {\tt{grd\_explr\_q=3}}} approaches the true theta relatively fast, while the learning strategy utilized by {\tt{grd\_splt}} is stuck in a theta that is further away from the true theta. However, the faster convergence to the true theta through exploration rounds comes at the cost of sub-par rewards collected during the exploration rounds. Recall that the exploration rounds in each epoch in general generate much smaller rewards compared to the exploitation rounds. This is manifested by the worse performance of {\tt{grd\_explr\_q=1}} compared to {\tt{grd\_explr\_q=2}} and {\tt{grd\_explr\_q=3}} which have fewer rounds of explorations and to {\tt{grd\_splt}} which has no exploration rounds. The fact that {\tt{grd\_splt}} performs on par to {\tt{grd\_kw}} indicates that there is indeed a region around the true theta that leads to similar optimal rewards, which further supports our Assumption \ref{assump:invariant_Approx_seed_set}.

\section{Conclusion}
In this paper, we propose a novel information diffusion model that is aware  of negative user opinions. Our linear threshold-based diffusion model allows users with different characteristics to have different information-sharing behaviors and attitude-formation patterns, while preserving monotonicity and submodularity. As a result, the greedy algorithm can be applied to achieve a $1-1/e - \epsilon$ approximation ratio for influence maximization. We further consider an online-learning problem in which the parameters of our diffusion model are initially unknown and need to be gradually learned through repeated rounds of negativity-aware influence maximization. Unlike existing works that assume the availability of edge-level observations, we conduct our analysis assuming only node-level feedback. We devise learning algorithms that achieve asymptotic expected average scaled regrets in the order of $\mathcal{O} (T^{-c})$, where $T$ is the number of rounds, $c < 1$ depends on the hyper-parameter input to our algorithms and can be arbitrarily close to 1. 

There are several interesting future research problems. For example, with our current model, after a node is activated, the effect of incoming positive weights and that of incoming negative weights are symmetric in determining the sign of the activation. We can also explore an asymmetric model, where the positive and negative influences are weighted differently. Another possible direction of research is to find a diffusion model that allows activated users to change their attitude, while preserving nice mathematical properties such as monotonicity and submodularity. 

\bibliographystyle{informs2014} 
\bibliography{references.bib}

\begin{appendices}
\section*{Online Appendix}
\section{IC-N with heterogeneous quality factors} \label{apd:IC-N-break}
\begin{figure}
    \centering
    \includegraphics[scale=0.3]{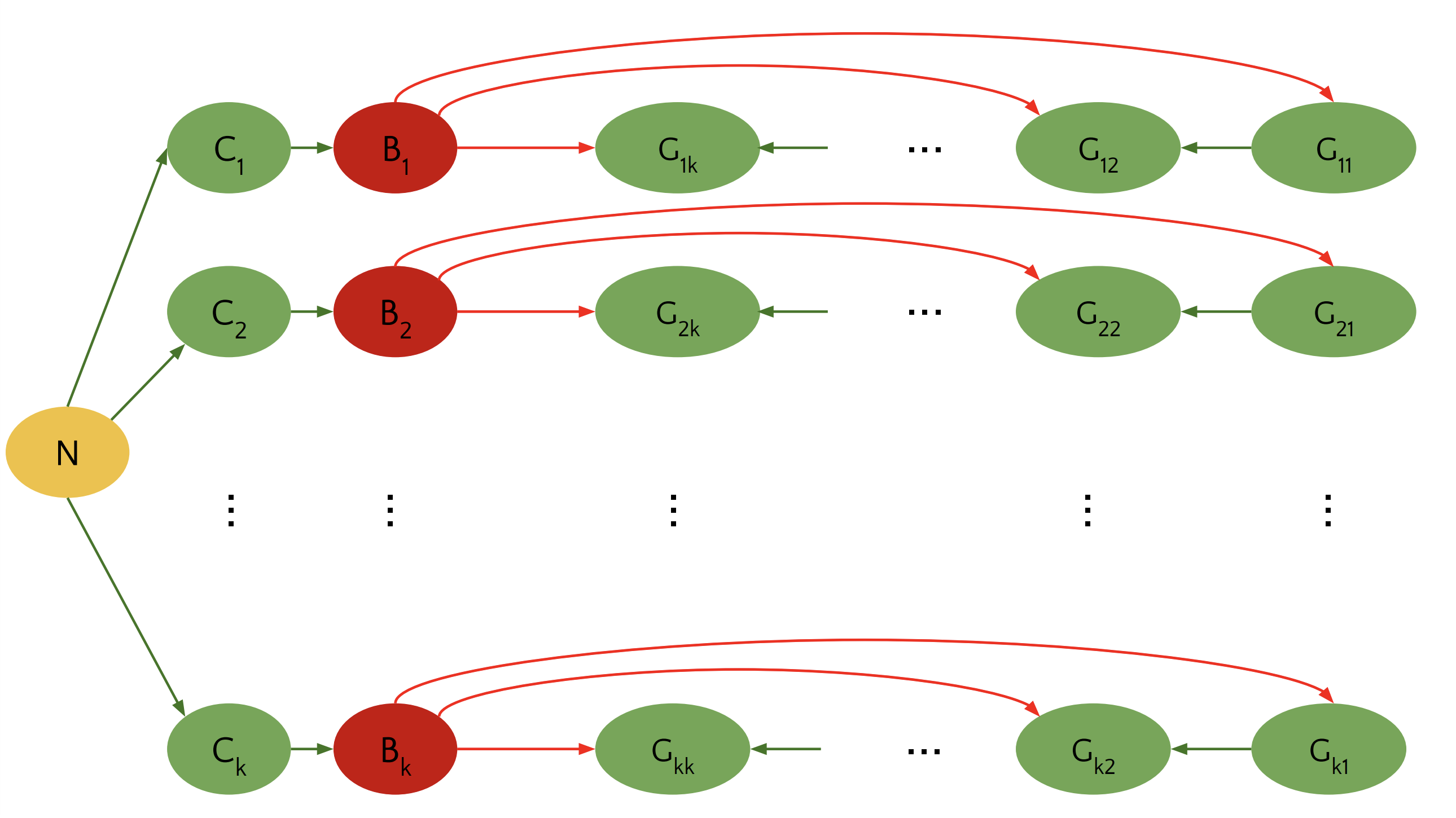}
    \caption{The case where weights and upper bounds cannot be estimated accurately by node-level observations.}
    \label{fig:1}
\end{figure}

Figure \ref{fig:1} illustrates an instance in which the greedy algorithm can have an arbitrarily bad approximation ratio when the quality factors are heterogeneous in the IC-N model proposed in \cite{chen2011influence}. In this instance, the influence probabilities $w(e)$'s equal to 1 for all $e \in \cE$. Nodes $B_i$'s for $i = 1, ..., k$ have quality factors $q(B_i) = 0$. All other nodes $v$ have quality factors $q(v)=1$. Assume the cardinality constraint on the seed set is a positive integer $k \in \mathbb{Z}^+$. For this specific instance, the optimal seed set is $S = \{G_{11}, G_{21}, ..., G_{k1}\}$ with an expected positive influence spread of $k \times k = k^2$. The first node the greedy algorithm picks, however, is $N$. This is because the expected positive influence after seeding $N$ is $k+1$ while seeding any other node generates an expected positive influence of at most $k$. As a result, the expected positive influence of the seed set chosen by greedy is at most that of the seed set $S' = \{N, G_{11}, G_{21}, ..., G_{(k-1)1}\}$. It can be easily calculated that the expected positive influence spread of $S'$ is $(k + 1) + 3\times(k-1) + (k-1)/2 = 4.5k - 2.5$, which can be arbitrarily worse than the optimal value $k^2$ as $k$ approaches infinity.

\section{Proof of Theorem \ref{thm:sig^+_mono_subm}} \label{sec:LT-N-mon-sub}

    
\subsection{Negativity-Aware Triggering Set model (TS-N)}\label{subsec:TS-N}
To prove the monotonicity and submodularity of $\sgplus(\cdot)$, we first define another diffusion model called \textit{Negativity-aware Triggering Set model} (TS-N). Let $\tilde{f}^+(S)$ be the expected number of positive nodes at the end TS-N with seed set $S$. We first prove the submodularity and monotonicity of  $\tilde{f}^+(\cdot)$. We then show that the distribution of number of positive nodes at the end of the TS-N process has the same distribution as that under LT-N. In this way, the submodularity and monotonicity of  $\tilde{f}^+(\cdot)$ allows us to conclude that $\sgplus(\cdot)$ have the same properties.

To mathematically describe the TS-N model, we first define an edge sampling process called Multinomial In-neighbor Sampling.

\begin{definition}[Multinomial in-neighbor sampling]\label{def:multinomial_X}
Consider a directed graph  $G(\cV,\cE)$  with edge weights $w: \cE \to [0,1]$ satisfying $\sum_{u\in \iN(v)} w(u,v)\leq 1$ where $\iN(v) = \{u: (u, v) \in \cE\}$. A \textit{multinomial in-neighbor sampling} $X: \cE \to \{0,1\}$ is defined as follows:
\begin{itemize}
    \item Each $v\in \cV$ independently chooses \textit{at most} one in-neighbor from $\iN(v)$ according to a multinomial distribution with probability $w(u,v)$ for each $u$ and $1-\sum_{u\in \iN(v)}w(u,v)$ for choosing nothing.
    \item For each $v$ that chooses an in-neighbor $u$, set $X(u,v)=1$. Set all other $X(e)=0$. 
\end{itemize}
Use $\cE_X$ to denote the set of chosen edges with respect to $X$ (called \textit{live edges}). Define the live-edge graph $\cG_X = (\cV,\cE_X)$.
\end{definition}

Note that each connected component of $G_X$ in Definition \ref{def:multinomial_X} is either a tree or a cycle because the degree of each node is at most 2.

Given a multinomial in-neighbor sampling $X$, we can further define a \emph{correction sampling} $Y_X$.
\begin{definition}[Correction Sampling]\label{def:sample_correction}
Consider a directed graph $G(\cV,\cE)$ with edge weights $w: \cE \to [0,1]$ satisfying $\sum_{u\in \iN(v)} w(u,v)\leq 1$ where $\iN(v) = \{u: (u, v) \in \cE\}$. Given a multinomial in-neighbor sampling $X: \cE \to \{0,1\}$, a correction sampling $Y_X: \{e: X(e) = 1\} \to \{-1, 0,1\}$ is defined as follows.
For each $(u,v) \in \cE$ such that $X(u,v) = 1$, sample a value from $\{-1, 0, 1\}$ according to a multinomial distribution with probabilities $q^-(v), 1-r(v), q^+(v)$, respectively.
\end{definition}

Given a seed set $S$, the Negativity-aware Triggering Set diffusion process is defined as follows.  
    \begin{enumerate}
    \item Obtain a multinomial in-neighbor sampling $X: \cE \to \{0,1\}$ of $\cG$ and the corresponding live-edge graph $\cG_X = (\cV, \cE_X)$ as defined in Definition \ref{def:multinomial_X}.
    \item Obtain a correction sampling $Y_X$ as defined in Definition \ref{def:sample_correction}.
    \item Now consider the following step-wise revelation of a multinomial in-neighbor sampling $X$ and the corresponding correction sampling $Y_X$: 
    
    \begin{itemize}
        \item Initiate $B_0(S) = B_0^+(S) = B_0^-(S) = \emptyset$.
        \item In step $1$, nodes in $S$ become positive. Let $B_1(S) = B_1^+(S) = S$, and $B_1^-(S) = \emptyset$. 
        \item For $\tau \geq 2$, initiate $B_\tau(S) = B_{\tau-1}(S)$, $B_\tau^+(S) = B_{\tau-1}^+(S)$, and $B_\tau^-(S) = B_{\tau-1}^-(S)$. Then, each node $v\in \cV\setminus B_{\tau-1}(S)$ with at least one in-neighbor in $B_{\tau-1}(S)$ samples its in-neighbor according to the multinomial distribution \textit{conditioned} on it being outside $B_{\tau-2}$(S). If an in-neighbor $u\in B_{\tau-1}(S)\setminus B_{\tau-2}(S)$ is sampled, add $v$ to $B_{\tau}(S)$ and reveal $Y_X(u,v)$. If $Y_X(u,v) = 0$, add $v$ to $B_{\tau}^+(S)$ if $u\in B_{\tau-1}^+(S)$ or to $B_{\tau}^-(S)$ if $u\in B_{\tau-1}^-(S)$. If $Y_X(u,v) = -1$, add $v$ to $B_{\tau}^-(S)$. If    
         $Y_X(u,v) = 1$, add $v$ to $B_{\tau}^+(S)$.
        \item The process terminates in round $T$ where $B_{T}(S) = B_{T-1}(S)$. Use $B(S) := B_{T}(S)$ to denote the set of active nodes at the end of the diffusion process. Denote by $B^+(S) := B^+_{T}(S)$  the set of positive nodes, and denote by  $B^-(S) := B^-_{T}(S)$ the set of negative nodes, with respect to a seed set $S$. Note that $B(S) = B^+(S) \cup B^-(S)$.
    \end{itemize}
    \end{enumerate}

    \subsection{Monotonicity and Submodularity of TS-N}

     Given seed set $S$ under a fixed multinomial in-neighbor sampling $X$ and correction sampling $Y_X$, we denote by  $f_{X, Y_X}^+(S) := |B^+(S)|$ the number of positively influenced nodes. By extending the idea of Lemma 4.4 in \cite{kempe2003maximizing}, we show that $f_{X, Y_X}^+(\cdot)$ is monotone  submodular in the following result. This implies that the expected influence functions $\tilde{f}^+(\cdot)$ is also monotone submodular.
    
    \begin{lemma}\label{prop:subm_f_X_TS-N}
    Let $X$ be a fixed multinomial in-neighbor sampling, and let $Y_X$ be a fixed correction sampling with respect to $X$. Under the TS-N model, $f^+_{X, Y_X}(\cdot)$ is monotone submodular. 
    \end{lemma}
    \proof{Proof.} 
    We first show that $f_{X,Y_X}^+(\cdot)$ is monotone. Let $S$ and $R$ be any two seed sets such that $S\subset R\subset \cV$. We want to show that $f_{X,Y_X}^+(S) \leq f_{X,Y_X}^+(R)$, i.e., $B^+(S) \subset B^+(R)$. We do so by proving $B_t^+(S) \subset B_t^+(R) \;\; \forall t \leq T$ using induction.
    
    First for $t = 0$, $B_0^+(S) = \emptyset = B_0^+(R)$. For $t = 1$, $B_1^+(S) = S \subset R = B_1^+(R)$.
    
    Now we show $B_t^+(S) \subset B_t^+(R)$ for $t \geq 2$. If $v \in B_t^+(S)$, then either $v \in B_{t-1}^+(S)$ or $v \in B_{t}^+(S) \backslash B_{t-1}^+(S)$. In the first case, by the inductive hypothesis, $v \in B_{t-1}^+(S) \subset B_{t-1}^+(R) \subset B_{t}^+(R)$. In the second case $v \in B_{t}^+(S) \backslash B_{t-1}^+(S)$, we can further consider three sub-cases:
    
    \begin{enumerate}
        \item $v \in R$: clearly $v \in B_{1}^+(R) \subset  B_{t}^+(R)$.
        \item $v \notin R$, in multinomial in-neighbor sampling $X$, $v$ samples a parent $u \in B_{t-1}^+(S) \backslash B_{t-2}^+(S)$, and $Y_X(u,v) = 0$: by inductive hypothesis, $u \in B_{t-1}^+(S) \backslash B_{t-2}^+(S) \subset B_{t-1}^+(S) \subset  B_{t-1}^+(R)$. Therefore, $u \in B_k^+(R)$ for some $k \in 1, ..., t-1$. As a result, we have $v \in B_t^+(R)$.
        \item $v \notin R$, in multinomial in-neighbor sampling $X$, $v$ samples a parent $u \in B_{t-1}(S) \backslash B_{t-2}(S)$, and $Y_X(u,v) = 1$: we know that $u \in B_{t-1}(S) \backslash B_{t-2}(S) \subset B_{t-1}(S) \subset B_{t-1}(R)$ (Note: it is trivial to show $B_\tau(\cdot)$ is monotone for each $\tau$). Therefore, $u \in B_k(R)$ for some $k \in 1, ..., t-1$. As a result, we obtain $v \in B_t^+(R)$ since the correction sampling $Y_X(u,v)=1$ dictates $v$'s sign in this case.
    \end{enumerate}
    
    To show that $f_{X, Y_X}^+(\cdot)$ is submodular, we consider any two seed sets $S$ and $R$ such that $S\subset R\subset \cV$ and any node $v \in \cV \backslash R$. It suffices to show \[B^+(R\cup \{v\})\setminus B^+(R) \subseteq B^+(S\cup \{v\})\setminus B^+(S).\] For each node $u \in B^+(R\cup \{v\})\setminus B^+(R)$, there must exist a path from $v$ to $u$ in $\cG_X$ that does not intersect with $R$. Meanwhile, the correction sampling $Y_X$ on the edges in the path results in $u$ is positive. As a result, $u$ contributes to $B^+(S \cup \{v\})$ too through this path. Since $u \notin B^+(R)$, by monotonocity of $B^+(\cdot)$ proved previously, we also have $u \notin B^+(S)$. Therefore, we conclude that $u \in B^+(S\cup \{v\})\setminus B^+(S)$ as well.
    \QED
    
    \begin{corollary}\label{coro:subm_f_TS-N}
    Under the TS-N diffusion process, the function $\tilde{f}^+(\cdot)$ is monotone submodular. 
    \end{corollary}
    \proof{Proof.} Let $\tilde{f}^+(\cdot) = \sum_{X,Y_X} \PP[X,Y_X] f^+_{X,Y_X}(\cdot)$, where the sum is taken over all possible multinomial in-neighbor sampling realizations $X$ and correction samplings $Y_X$. 
    From Lemma \ref{prop:subm_f_X_TS-N}, for any $X,Y_X$, one can see $f^+_{X,Y_X}(\cdot)$ is monotone submodular. 
    Since a non-negative linear combination of monotone submodular functions is also monotone submodular, we conclude that $\tilde{f}^+(\cdot)$ is monotone submodular. \QED 
    
\subsection{Monotonicity and Submodularity of LT-N}
    Recall that $A(S)$ (resp. $A^+(S)$, $A^-(S)$) denotes the set of active (resp. positive, negative) nodes at the end of a LT-N diffusion process, with $S$ being the initial seed set. The randomness of $A(S)$, $A^+(S)$ and $A^-(S)$ comes from the random thresholds $\{b_v : v\in \cV \}$. Let $B(S)$ (resp. $B^+(S)$, $B^-(S)$) denote the set of active (resp. positive, negative) nodes at the end of a TS-N diffusion process, with $S$ being the initial seed set. The randomness of $B(S)$, $B^+(S)$ and $B^-(S)$ comes from the random multinomial in-neighbor sampling $X$ and correction sampling $Y_X$. In this section, we show that $A(S)$ (resp. $A^+(S)$, $A^-(S)$) and $B(S)$ (resp. $B^+(S)$, $B^-(S)$) have the same distribution for any given seed set $S$. Then with the distributional equivalence, we derive the monotonicity and submodularity results for LT-N. Our result is a generalization of Theorem 4.6 in \cite{kempe2003maximizing}.
    
    In the description of the LT-N process, $A_{\tau}(S)$ (resp. $A_{\tau}^+(S)$, $A_{\tau}^-(S)$) are used to denote the set of active (resp. positive, negative) nodes by the end of step $\tau$ given seed set $S$. For the remainder of this section, we simplify the notation by omitting the dependency of $A_{\tau}(S)$, $A_{\tau}^+(S)$, and $A_{\tau}^-(S)$ on $S$. We define $\cH_\tau^A := (A_{t}, A_{t}^+, A_{t}^-)_{t=0}^{\tau}$ as the history of the $A_t$, $A_t^+$, and $A_t^-$'s for $t = 0, ..., \tau$.
    
    Similar to LT-N, we define $\cH_\tau^B := (B_{t}, B_{t}^+, B_{t}^-)_{t=0}^{\tau}$ as the history of the $B_t$, $B_t^+$, and $B_t^-$'s for $t = 0, ..., \tau$ in the TS-N process (given the same seed set $S$, which is omitted in the notation). We first prove the following result on the equivalence in distribution of $\cH_\tau^A$ and $\cH_\tau^B$.
    
    \begin{lemma}\label{prop:distri_equiv}
    For any $\tau \geq 0$, $\cH_\tau^A \eqd \cH_\tau^B$.
    \end{lemma}
    
    \proof{Proof.} 
    In LT-N, nodes in $S$ become positive in step $1$ ($A_1 = A_1^+ = S$ and $A_1^-=\emptyset$).
    In step $\tau \geq 2$, the conditional probabilities that each node $v\in \cV\setminus A_{\tau-1}$ becomes active, positive, and negative, given the history $\cH_{\tau-1}^A$, are as follows: 
    \begin{align}
        \PP[v\in A_{\tau} |\cH_{\tau-1}^A] & = \frac{\sum_{u\in \iN(v)\cap(A_{\tau-1}\setminus A_{\tau-2})} w(u,v)}{1 - \sum_{u\in \iN(v)\cap A_{\tau-2}}w(u,v)}, \label{p_Atau_cond} \\[1em]
        \PP[v\in A^+_{\tau} | \cH_{\tau-1}^A] & = \frac{\sum_{u \in \iN(v) \cap (A_{\tau-1}\setminus A_{\tau-2})} w(u, v)}{1 - \sum_{u\in \iN(v)\cap A_{\tau-2} }w(u,v)} \\
        & \ \cdot \left(q^+(v) + [1-r(v)] \frac{\sum_{u \in \iN(v) \cap (A^+_{\tau-1}\setminus A_{\tau-2})} w(u, v)}{\sum_{u \in \iN(v) \cap (A_{\tau-1}\setminus A_{\tau-2})} w(u, v)} \right),\label{p_Atau+_cond} \\[1em]
        \PP[v\in A^-_{\tau} |\cH_{\tau-1}^A] &= \frac{\sum_{u \in \iN(v) \cap (A_{\tau-1}\setminus A_{\tau-2})} w(u, v)}{1 - \sum_{u\in \iN(v)\cap A_{\tau-2}} w(u,v)} \\
        & \ \cdot \left(q^-(v) + [1-r(v)] \frac{\sum_{u \in \iN(v) \cap (A^-_{\tau-1}\setminus A_{\tau-2})} w(u, v)}{\sum_{u \in \iN(v) \cap (A_{\tau-1}\setminus A_{\tau-2})} w(u, v)} \right).\label{p_Atau-_cond}
    \end{align}
    
    Now we consider the TS-N process with the step-wise revealing of the sample edges in $X$. In step $1$, nodes in $S$ become positive ($B_1 = B_1^+ = S$ and $B_1^- = \emptyset$).
    
     In step $\tau \geq 2$, for each $v\in \cV \setminus B_{\tau-1}$, the conditional probabilities of $v$ being active in this step given the history $\cH_{\tau-1}^B$ are as follows:
    \begin{align}
        \PP[v\in B_{\tau}|\cH_{\tau-1}^B]
        &=\frac{\PP[X\ \text{chooses an in-neighbor from } B_{\tau-1}\setminus B_{\tau-2}]}{\PP[X\ \text{does not choose any in-neighbor from} \ B_{\tau-2}]}\nonumber \\
        &= \frac{\sum_{u\in \iN(v)\cap  (B_{\tau-1}\setminus B_{\tau-2})} w(u,v)}{1 - \sum_{u\in \iN(v) \cap B_{\tau-2}} w(u,v)}, \label{p_Btau_cond} \\[1em]
        \PP[v\in B^+_{\tau}|\cH_{\tau-1}^B] 
        &=\frac{\PP[X\ \text{chooses an in-neighbor } u \in B_{\tau-1}\setminus B_{\tau-2}]}{\PP[X\ \text{does not choose any in-neighbor from}\ B_{\tau-2}]} \\
        &\ \cdot \PP[\{u \in B^+_{\tau-1} \backslash B^+_{\tau-2}, Y_X(u,v) = 0\} \ \text{or} \ Y_X(u,v) = 1]\nonumber \\
        &= \frac{\sum_{u \in \iN(v) \cap (B^+_{\tau-1}\setminus B^+_{\tau-2})} w(u, v)}{1 - \sum_{u\in \iN(v)\cap B_{\tau-2} }w(u,v)} \\
        & \  \cdot \left(q^+(v) + [1-r(v)] \frac{\sum_{u \in \iN(v) \cap (B^+_{\tau-1}\setminus B_{\tau-2})} w(u, v)}{\sum_{u \in \iN(v) \cap (B_{\tau-1}\setminus B_{\tau-2})} w(u, v)} \right),\label{p_Btau+_cond}\\[1em]
        \PP[v\in B^-_{\tau} | \cH_{\tau-1}^B] 
        &=\frac{\PP[X\ \text{chooses an in-neighbor from } B^-_{\tau-1}\setminus B_{\tau-2}]}{\PP[X\ \text{does not choose any in-neighbor from}\ B_{\tau-2}]}\nonumber \\
        &\ \cdot \PP[\{u \in B^-_{\tau-1} \backslash B^-_{\tau-2}, Y_X(u,v) = 0\} \ \text{or} \ Y_X(u,v) = -1]\nonumber \\
        &= \frac{\sum_{u \in \iN(v) \cap (B^+_{\tau-1}\setminus B^+_{\tau-2})} w(u, v)}{1 - \sum_{u\in \iN(v)\cap B_{\tau-2} }w(u,v)} \\
        &\ \cdot \left(q^-(v) + [1-r(v)] \frac{\sum_{u \in \iN(v) \cap (B^-_{\tau-1}\setminus B_{\tau-2})} w(u, v)}{\sum_{u \in \iN(v) \cap (B_{\tau-1}\setminus B_{\tau-2})} w(u, v)} \right).\label{p_Btau-_cond}
    \end{align} 
    
    We now prove the statement of the lemma by induction on $\tau$. For $\tau = 0$, the lemma holds since $A_0=B_0=A_0^+=B_0^+=A_0^-=B_0^- = \emptyset$. For $\tau = 1$, the lemma holds because $A_1=B_1=A_1^+=B_1^+=S$, and $A_1^-=B_1^- = \emptyset$. 
    As the inductive hypothesis, we assume that $\cH_{\tau-1}^A \eqd \cH_{\tau-1}^B$ for some $\tau\geq 2$. Let $H_{\tau-1}$ be any realization of $\cH_{\tau-1}^A$. By induction hypothesis, we have  \begin{equation}
    \PP(\cH_{\tau-1}^A = H_{\tau-1}) = \PP(\cH_{\tau-1}^B = H_{\tau-1}).\label{eq:ind_hyp}
    \end{equation} 
    Furthermore, from \eqref{p_Atau_cond} and \eqref{p_Btau_cond}, we have that \begin{align}
    \PP[v\in A_{\tau}|\cH_{\tau-1}^A = H_{\tau-1}] = \PP[v\in B_{\tau}|\cH_{\tau-1}^B = H_{\tau-1}].\label{induc_A_eq_B}
    \end{align}
    
    Similarly, from \eqref{p_Atau+_cond} and \eqref{p_Btau+_cond}, we have that 
    \begin{align}
        \PP[v\in A_{\tau}^+|\cH_{\tau-1}^A = H_{\tau-1}] = \PP[v\in B_{\tau}^+ |\cH_{\tau-1}^B = H_{\tau-1}].\label{induc_A+_eq_B+}
    \end{align}
    
    Finally, from \eqref{p_Atau-_cond} and \eqref{p_Btau-_cond}, we get
    \begin{align}
        \PP[v\in A_{\tau}^-|\cH_{\tau-1}^A = H_{\tau-1}] = \PP[v\in B_{\tau}^- | \cH_{\tau-1}^B = H_{\tau-1}]. \label{induc_A-_eq_B-}
    \end{align}
    
    From \eqref{induc_A_eq_B}-\eqref{induc_A-_eq_B-}, we have that the following conditional distributions are equal to each other: 
    \begin{equation}
    ((A_{\tau}, A_{\tau}^+, A_{\tau}^-)|\cH_{\tau-1}^A = H_{\tau-1}) \eqd \ ((B_{\tau}, B_{\tau}^+, B_{\tau}^-)|\cH_{\tau-1}^B = H_{\tau-1}).\label{eq:cond_equiv}\end{equation}
   Using \eqref{eq:ind_hyp} and \eqref{eq:cond_equiv}, we obtain $\cH_{\tau}^A \eqd \cH_{\tau}^B$ as desired.
    
For any seed set $S\subseteq \cV$, Lemma \ref{prop:distri_equiv} implies $A^+(S) \eqd B^+(S)$. Therefore, we have $\sgplus(S) = \EE[|A^+(S)|] = \EE[|B^+(S)|] = \tilde{f}^+(S)$. From Corollary \ref{coro:subm_f_TS-N}, we have that $\tilde{f}^+(\cdot)$ is monotone submodular. Therefore, we can conclude that $f^+(\cdot)$ is as well.\QED

\section{Proofs for Online Learning Problems} \label{apd:online-lemmas}

\subsection{Proof of Lemma \ref{thm:2}}\label{apd:online-lemmas-thm:2}
\proof{Proof of Lemma \ref{thm:2}.}
Consider the set of $\alpha$-approximation seed sets $\cA(\theta^*,\alpha)$. Let $g(S,w_{\theta^*}) := f(S,w_{\theta^*})/f^{\text{opt}}(w_{\theta^*})$. We call $g(S,w_{\theta^*})$ the \textit{approximation factor} of seed set $S$. By definition, we have $g(S,w_{\theta^*}) \geq \alpha$ for any $S \in  \cA(\theta^*,\alpha)$. We prove that a sufficient condition for Assumption \ref{assump:invariant_Approx_seed_set} to hold is $\min_{S \in \cA(\theta^*,\alpha)} g(S,w_{\theta^*})> \alpha$.

Given any seed set $S$ of size $K$, by Lemma \ref{lemma:continuity} in Appendix Section \ref{apd:aux-lemmas-continuity}, it is easy to see that $f(S,w_\theta)$ and $f^{\text{opt}}(w_{\theta})$ are continuous in $\theta$, which implies that $g(S,w_{\theta})$ is also continuous in $\theta$. Suppose  $\min_{S \in \cA(\theta^*,\alpha)} g(S,w_{\theta^*})> \alpha$, i.e., $g(S,w_{\theta^*})> \alpha$ for all $S \in \cA(\theta^*,\alpha)$. For any $S \in \cA(\theta^*,\alpha)$, since $g(S,w_\theta)$ is continuous,  there exists a positive constant $\delta_S >0$ such that $g(S,w_\theta) > \alpha$ for all $\| \theta - \theta^*\|\leq \delta_S$. Let $\delta_1 = \min_{S \in \cA(\theta^*,\alpha)} \delta_S$.  We conclude that  $g(S,w_\theta) >\alpha $ for all $S \in \cA(\theta^*,\alpha)$ and $\| \theta - \theta^*\| \leq \delta_1$.  This is equivalent to  $\cA(\theta^*,\alpha)  \subset \cA(\theta,\alpha)$ for $\|\theta - \theta^* \| \leq \delta_1$. 

At the same time, consider the seed sets that are not $\alpha$-approximations to $f^{\text{opt}}(w_{\theta^*})$, i.e.,  $g(S,w_{\theta^*}) < \alpha$ and thus $S \notin \cA(\theta^*,\alpha)$. By a similar argument, we conclude that there exists a positive constant $\delta_2 >0$ such that for all $S \notin \cA(
\theta^*,\alpha)$, $g(S,w_\theta) < \alpha$ for all $\|\theta - \theta^* \|\leq \delta_2$, which further implies $\cA(\theta,\alpha) \subset \cA(\theta^*,\alpha)$.
By choosing $\delta_0 = \min(\delta_1,\delta_2)$, we conclude $\cA(\theta,\alpha) = \cA(\theta^*,\alpha)$ for all $\|\theta - \theta^* \|\leq \delta_0$.

The preceding analysis shows that Assumption \ref{assump:invariant_Approx_seed_set} indeed holds if $\min_{S \in \cA(\theta^*,\alpha)} g(S,w_{\theta^*})> \alpha$, which yields a contradiction. We thus conclude that Assumption \ref{assump:invariant_Approx_seed_set} holds whenever $\min_{S \in \cA(\theta^*,\alpha)} g(S,w_{\theta^*})> \alpha$, and fails  
only when there exists a set $S$ such that $f(S,w_{\theta^*}) = \alpha \cdot f^{\text{opt}}(w_{\theta^*})$.

Finally, consider an LT-N network with fixed $w_{\theta^\ast}$. There are finitely many seed sets and their corresponding approximation factors are distributed as finite discrete values on $[0,1]$. Since $\alpha = 1-1/e - \epsilon$ is required to be decided prior to calling the $(\alpha,\gamma)$-approximation oracle, if we sample $\alpha$ uniformly over an interval $[1-1/e-\epsilon_1, 1-1/e-\epsilon_2]$ with $\epsilon_1 > \epsilon_2$, the probability of $\alpha$ being equal to one of the approximation factors is zero. Thus, we conclude that $f(S,w_{\theta^*}) = \alpha\cdot f^{\text{opt}}(w_{\theta^*})$ is a zero-probability event, which completes the proof. 
\QED

\subsection{Proof of Theorem \ref{thm:IM01}}\label{apd:online-lemmas-IM01}
\proof{Proof of Theorem \ref{thm:IM01}:} Let $f^* = f(S^{\text{opt}}, w^*)$ be the optimal expected influence in any round under the constraint that the cardinality seed set $S$ cannot exceed $K$. Define $R_t^{\alpha \gamma} = f^* - \EE[\frac{1}{\alpha \gamma} f(S_t,w^*)] $  as the expected $(\alpha,\gamma)$-scaled regret in round $t$, where $S_t$ is the seed set that the algorithm selects in round $t$. The expectation is taken over the randomness of $S_t$ in an exploitation round, and reduces to a deterministic function in an exploration round, as we have pre-fixed the choice of exploration nodes. Note that the randomness in the diffusion process has already been captured in the definition of $f$.

We denote by $k_t$ the index of the epoch that round $t$ is in. For an exploration round $t \leq T$, Algorithm~\ref{alg:IM02} seeds a single node, which yields a regret $R_t^{\alpha \gamma} \leq f^*$. For an exploitation round $t\leq T$, Algorithm~\ref{alg:IM02} uses the least-squared estimate $\theta_{k_t} = \M_{k_t}^{-1} r_{k_t}$ to estimate edge weights. 


Let $D$ be a known upper bound on $\| \theta^*\|$, and let $c_k = \sqrt{d \log  (1+ kd )  + d\log(k^q )}   + D $, we define the confidence region $\cC_k$ as 
\begin{equation*}
   \cC_{k} :=\Big \{ \theta:   \| \theta - \theta_{k}\|_{\M_{k}} \leq c_k\Big \}. 
\end{equation*}
In epoch $k$, we define the favorable event $\xi_{k}$ as
\begin{equation}
    \xi_{k}:=  \I \left\{   \theta^* \in \cC_k \right\}, \label{good_event}
\end{equation}
and $\bar \xi_{k}$ as the complement of $\xi_{k}$. By using Theorem 2 of \cite{abbasi2011improved}, we have $\PP(\xi_k) \geq 1-1/k^q$. 

We then consider the regret incurred within epoch $k$. 
First, as epoch $k$ includes $d$ exploration rounds, a fixed regret of size at most $f^* d$ is always incurred in exploration. Next, for any exploitation round $t = t_k+d,\cdots, t_{k+1}-1$ belonging to epoch $k$,
given $\xi_{k}$, the algorithm yields a regret $\EE[R_t^{\alpha \gamma}|\xi_{k}]$ when the favorable event holds, and a regret $f^*$ if the unfavorable event $\bar \xi_k$ happens. In summary, we have 
\begin{equation} \label{eq:total_regret_LT_01}
\begin{split}
\EE[R_t^{\alpha \gamma}]   \leq 
\PP(\xi_{k})  \EE[  R_t^{\alpha \gamma} | \xi_{k}] + \PP(\bar \xi_{k}) f^* = 
\PP(\xi_{k})  \EE[  f(S^{\text{opt}},w^*) - \frac{1}{\alpha \gamma} f(S_t,w^*) | \xi_{k}] + \PP(\bar \xi_{k}) f^*, 
\end{split}
\end{equation}
in an exploitation round $t$, where the equality comes from the definition of $(\alpha,\gamma)$-scaled regret.

Let $\epsilon > 0$ be the stability parameter in Assumption \ref{assump:invariant_Approx_seed_set} such that $\cA(\theta, \alpha) = \cA( \theta^*, \alpha) $  for all $\|\theta - \theta^* \| \leq \epsilon$. Clearly, $\|\theta - \theta^* \| \leq \epsilon$ serves as a sufficiently condition for $\cA(\theta, \alpha) = \cA( \theta^*, \alpha) $. We then investigate the performance of exploitation rounds based on whether $\| \theta_k - \theta^* \| \leq \epsilon$ holds.

Suppose $\| \theta_k - \theta^* \| > \epsilon$, the total exploitation regret within this epoch is upper bounded by $k^q f^*$.

Suppose on the other hand that $\|\theta_k - \theta^* \| \leq \epsilon$, which implies $\cA(\theta_k, \alpha) = \cA( \theta^*, \alpha)$. In this case, as the estimate $ \theta_k$  falls in the stable region, for exploitation rounds $t= t_k + d, \cdots, t_{k+1} -1$, we have 
\begin{equation}\label{eq:reg_good}
\begin{split}
     & \EE[R_t^{\alpha \gamma}|\xi_k] = \EE[  f^* - \frac{1}{\alpha \gamma} f(S_{t},w^*)|\xi_k] \\
   & \quad \leq  \EE \Big [  f^* - \frac{1}{\alpha \gamma} (\gamma \cdot \alpha f^* + (1-\gamma) \cdot 0) \Big ] \\
 & \quad =  \EE\left[  f^* - \frac{1}{\alpha \gamma} \cdot\gamma\alpha f^*\right] = 0.
\end{split}
\end{equation}
Combining \eqref{eq:total_regret_LT_01} and \eqref{eq:reg_good}, we have that given $\|\theta_k - \theta^* \| \leq \epsilon$, the total expected scaled regret incurred during the $k^q$ exploitation rounds in epoch $k$, namely, $t= t_k+d,\cdots, t_{k+1}-1$ is 
\begin{equation}\label{reg:stable}
    \sum_{t=t_k+d}^{t_{k+1}-1} \EE[R_t^{\alpha \gamma}] = \sum_{t=t_k+d-1}^{t_{k+1}-1} \PP(\xi_k) \EE[R_t^{\alpha \gamma}|\xi_k] + \sum_{t=t_k+d-1}^{t_{k+1}-1} \PP(\bar \xi_k) f^* \leq  f^*.
\end{equation}
Combining the above with the $f^*d$ regret incurred in $d$ exploration rounds, the regret incurred in epoch $k$ is upper bounded by $(d+1)f^*$ if $\|\theta_k - \theta^* \| \leq \epsilon$, and is upper bounded by $(d+k^q)f^*$ otherwise.

We now study the sufficient condition for $\|\theta_k - \theta^* \| \leq \epsilon$.
Let $\lambda_{\min}(\M_{k})$ be the smallest eigenvalue of $\M_{k}$ and recall $\lambda^\circ_{\min}$ is the smallest eigenvalue of the feature covariance matrix of the exploration nodes defined in Assumption \ref{assump:exploration_nodes}. 
The exploration rounds in Algorithm \ref{alg:IM02} guarantee the addition of $\sum_{e\in \cD^\circ} x_e x_e^T$ to the cumulative covariance matrix $\M_{k}$ in every epoch. We use the following lemma to get a lower bound on the increase in $\lambda_{\min}(\M_{k})$ after every epoch. 
\begin{lemma}\label{lemma:eig}[\cite{wilkinson1965algebraic}]
Let $M$ and $E$ be $d \times d$ symmetric matrices and denote $\lambda_{i}(A)$ as the $i$-th biggest eigenvalue of  any matrix $A$. Then
\[
\lambda_{i}(M) + \lambda_d(E) \leq \lambda_{i}(M+E) \leq \lambda_{i}(M) + \lambda_1 (E), \,\,\, \forall i =1,\ldots, d.
\]
\end{lemma}

If we let $M = \M_{k}$ and $E =\sum_{e\in \cD^\circ} x_e x_e^T$, then when $i = d$, we have by the first inequality in Lemma \ref{lemma:eig} that $\lambda_{\min}(\M_{k}) \geq \lambda_{\min} ( \M_{k-1}) + \lambda_{\min}^\circ$. By telescoping, we have $\lambda_{\min}(\M_{k}) \geq \lambda_{\min}^\circ k$.


Since $\lambda_{\min}(\M_{k}) \geq \lambda_{\min}^\circ k$, if $\theta^* \in \cC_k$, then we have 
\begin{equation*}\|\theta_k - \theta^*\|^2 \leq \frac{c_k^2} {\lambda_{\min}^\circ k}.
\label{eq:theta_bound}\end{equation*} 

Therefore, $\epsilon^2 \geq \frac{c_k^2}{\lambda_{\min}^{0} k }$ is  a sufficient condition for 
 $\| \theta_k - \theta^* \| \leq \epsilon $ to be satisfied. 

Recall that $c_k = \sqrt{d \log  (1+ kd )  + d\log(k^q )}   + D $. As a result, $g(k) := \frac{c_{k}^2} {\lambda_{\min}^\circ k} \to 0$ as  $k \to \infty$. Let $k_0$ be the smallest integer such that $g(k) \leq \epsilon^2$ for all $k \geq k_0$. Clearly, for epoch $k\geq k_0$, the condition $\| \theta_k - \theta^* \| \leq \epsilon $ is always satisfied.

Thus far we have investigated the regret incurred in each epoch based on whether the condition $\| \theta_k - \theta^* \| \leq \epsilon $ is satisfied. 
We now conduct regret analysis over $T$ rounds using the preceding results established on individual epochs. 
    
    Recall that $t_k$ denotes the index of the first round of epoch $k$. We first consider the case when $T \leq t_{k_0}$. Clearly, the cumulative scaled regret could be upper bounded by 
    \begin{equation}\label{eq:bad}
    \sum_{t=1}^T \EE[R_t^{\alpha \beta}]\leq Tf^*.
    \end{equation}

    Then, we consider the case when $T\geq t_{k_0}$. Let ${k_T}$ be the epoch such that round $T$ falls in. In particular, we have $t_{k_T} \leq T \leq t_{k_T} - 1$.

    We could further split the cumulative regret into the regret incurred before epoch $k_0$, the regret incurred during epochs $k_0,k_0+1,\cdots, k_T-1$, and the regret incurred during epoch $k_T$. 
    
    Based on our preceding analysis, the regret incurred in epoch $1,2,\cdots, k_0-1$ is upper bounded by $(t_{k_0} -1) f^*$. Meanwhile, for epochs $k_0,k_0+1,\cdots, k_T-1$, the regret incurred in each of them could be upper bounded by $(d+1)f^*$ based on \eqref{reg:stable}. It suffices to consider the regret incurred in epoch $k_T$ to complete our analysis. 
    
    Consider epoch $k_T$. Suppose $T \leq t_{k_T}+d$, which implies that round $T$ is in the exploration phase of epoch $k_T$. In this case, the cumulative regret incurred within the epoch is $(T-t_{k_T}) f^*$. Combining with previous analysis, the cumulative regret could be then expressed as 
    \begin{equation}
        \sum_{t=1}^T \EE[R_t^{\alpha \beta}] \leq (t_{k_0} -1) f^* + (k_T-k_0)(d+1)f^* + (T-t_{k_T}) f^*.
    \end{equation}
    
    Now suppose $t_{k_T} + d < T$, which implies that round $T$ is in the exploitation phase of epoch $k_T$, one can see that regret of size $f^*d$ is  incurred in exploration phase of epoch $k_T$. Meanwhile, using the fact that $\PP(\bar \xi_{k_T}) \leq 1/k_T^q$, the regret incurred within the exploration phase is 
    $$
    \sum_{t=t_{k_T} + d }^T \EE[R_t^{\alpha \beta}] \leq (T-t_{k_T} -d)f^*/k_T^q.
    $$
    Using similar analysis, the cumulative regret could be expressed as 
        \begin{equation}
        \sum_{t=1}^T \EE[R_t^{\alpha \gamma}] \leq (t_{k_0} -1) f^* + (k_T-k_0)(d+1)f^* + d f^* + \frac{(T-t_{k_T} - d)}{k_T^q}f^*.
    \end{equation}

Note that as $T-t_{k_T} - d \leq k_T^q$, the regret incurred in cases $T \leq t_{k_T}+d$ and $t > t_{k_T}+d$ can both be upper bounded by 
 \begin{equation}\label{eq:good}
        \sum_{t=1}^T\EE[R_t^{\alpha \gamma}] \leq (t_{k_0} -1) f^* + (k_T-k_0+1)(d+1)f^* .
    \end{equation}

Combining \eqref{eq:bad} and \eqref{eq:good}, we conclude that the cumulative regret of Algorithm \ref{alg:IM02} is bounded by 
\begin{equation*}
\sum_{t=1}^T \EE[R_t^{\alpha \gamma}] \leq 
\begin{cases}
  T f^* & \text{ if } T < t_{k_0},\\
\Big ( (t_{k_0} -1)  + (k_T-k_0+1)(d+1) \Big ) f^*& \text{ if }   T \geq t_{k_0}, 
    \end{cases}
\end{equation*}
which completes the proof.  
\QED

\subsection{Proof of Theorem \ref{thm:IM03}}\label{apd:online-lemmas-IM03}
\proof{Proof of Theorem \ref{thm:IM03}.}
Suppose Algorithm \ref{alg:IM03} has run $k$ epochs.
Let $D_{\beta}$ be a known upper bound on $\| \beta^*\|$, and let $$
\kappa_k = \sqrt{d' \log \left(1+ kd' \right) + d' \log (k^q) }  + D_{\beta}.
$$
Define the favorable event $\zeta_{k}$ as 
\begin{equation*}
    \zeta_{k}:= \ind  \left\{(\beta_k - \beta^*)^\top \V_k (\beta_k - \beta^*) \leq \kappa_k^2\,\right\}, 
\end{equation*}
and $\bar \zeta_{k}$ as the complement of $\zeta_{k}$. By Theorem 2 of \cite{abbasi2011improved}, we have $\PP(\zeta_k) \geq 1-1/k^{q}$.

Denote by $w_{\min}^{auto} :=  \min _{1\leq i \leq d'} w_i^{auto}$ the minimal incoming weight sum over all nodes in the autonomy exploration set $\mathcal D^{auto}$. At the end of $k$-th epoch, we consider a more strict favorable event:  $$\psi_k: = \xi_k \cap \zeta_k \cap \ind \{ \lambda_{\min}(\V_k) \geq ( k\cdot   w_{\min}^{auto}     - \varepsilon_k \big )  \}.$$
We define $\bar \psi_k$ as its complement.
By Lemma \ref{lemma:min_eig_perturb} in Appendix Section \ref{apd:online-lemmas-min_eig_perturb}, we have 
$\lambda_{\min}(\V_k) \geq ( k\cdot   w_{\min}^{auto}     - \varepsilon_k \big )  \cdot \lambda_{\min}^{auto}$ with probability at least $1- d'/k^{q}$. 
Under $\psi_k$, we have  $\| \theta_k - \theta^*\|_{\M_k}\leq c_k$ and $\| \beta_k - \beta^*\|_{\V_k}\leq \kappa_k$. 
Then,
by union of probability, it is clear that 
$
\P(\psi_k) \geq 1- (d+2d'+1)/k^{q} $ and
$
 \P( \bar \psi_k) \leq (d+2d'+1)/k^{q}. 
$

Let $\epsilon_{\theta}$ and $\epsilon_{\beta}$ be the stability parameters defined in Assumption \ref{assump:invariant_Approx_seed_set_LTN}, the corresponding stable regions are $\{\theta: \| \theta - \theta^* \| \leq \epsilon_{\theta}\}$ and $\{\beta: \| \beta - \beta^* \| \leq \epsilon_{\beta} \}$. Under $\psi_k$, a sufficient condition for 
our estimate $\theta_k$ and $\beta_k$ to fall in the stable region is $ \{ \frac{c_k^2}{\lambda_{\min}^o k} \leq \epsilon_\theta^2,  \frac{\kappa_k^2}{k\cdot   w_{\min}^{auto}     - \varepsilon_k } \leq \epsilon_\beta^2  \}$. Similar as in the proof of Theorem \ref{thm:IM01}, the functions $g(k) := \frac{c_k^2}{\lambda_{\min}^o k} $ and $h( k) :=  \frac{\kappa_k^2}{k\cdot   w_{\min}^{auto}     - \varepsilon_k }$ monotonically decrease to 0 as $k\to \infty$.

Let $k_0 = \min\{k :\frac{c_k^2}{\lambda_{\min}^o k} \leq \epsilon_\theta^2,  \frac{\kappa_k^2}{k\cdot   w_{\min}^{auto}     - \varepsilon_k } \leq \epsilon_\beta^2\}$ be the minimal number of epochs required to achieve the above condition, and let $t_{k_0}$ be the corresponding round index. Then, for any epoch $k \geq k_0$ with rounds $t= t_{k},\cdots, t_{k+1}-1$, using similar analysis as in the proof of Theorem \ref{thm:IM01}, the regret incurred within this epoch is bounded by 
\begin{equation} \label{eq:good_regret_LTN}
    \sum_{t=t_k}^{t_{k+1}-1} \EE[R_t^{\alpha \gamma}] \leq  (d+ 2d'+1)f^*.
\end{equation}
Now we provide the finite regret bounds based on whether $T \leq t_{k_0}-1$ or not. 
Suppose $T \leq t_{k_0}-1$, it is clear that 
\begin{equation}\label{eq:bad_LTN}
    \sum_{t=1}^T \EE[R_t^{\alpha \gamma}] \leq T f^*.
\end{equation}
On the other hand, suppose $T \geq t_{k_0}$, using similar analysis as in the proof of Theorem \ref{thm:IM01}, together with \eqref{eq:good_regret_LTN} that the regret incurred within each epoch $k\geq k_0$ is upper bounded by $(d+2d'+1)f^*$, the cumulative regret could be bounded by 
\begin{equation}\label{eq:good_LTN}
  \sum_{t=1}^T \EE[R_t^{\alpha \gamma}] \leq   \Big ( (t_{k_0} -1)  + (k_T-k_0)(d+ 2d'+ 1) \Big ) .
\end{equation}
Combining \eqref{eq:bad_LTN} and \eqref{eq:good_LTN}, we conclude  the proof. 
\QED

\subsection{Lemma \ref{lemma:min_eig_perturb} and Its Proof} \label{apd:online-lemmas-min_eig_perturb}
\begin{lemma} \label{lemma:min_eig_perturb}
Suppose Algorithm \ref{alg:IM03} has run $k$ epochs, let $w_{\min}^{auto} = \min_{1\leq i \leq d'} w_i^{auto}$, and let $\varepsilon_k = \sqrt{\frac{q \log k }{2k}}$, then with probability at least $1- d'/k^{q}$, we have 
$$
\lambda_{\min}(\V_k) \geq \big ( k\cdot   w_{\min}^{auto}     - \varepsilon_k \big )  \lambda_{\min}^{auto} .
$$
 \end{lemma}
\proof{Proof.}
Consider the $d'$ exploration nodes $v_i^{auto}$, for $i = 1, ..., d'$ that are used to update the belief on $\beta^*$. For each $v_i^{auto}$, let $S_i^{auto}$ be the set of parent nodes of $v_i^{auto}$ with the smallest indices such that $|S_i^{auto}| = \min\{|\cN_{in}(v_i^{auto})|, K\}$. Recall that in the $i$-th exploration round of autonomy factor in each epoch $m$, we choose $S_i^{auto}$ as the seed set; with probability $w_i^{auto} := \sum_{u \in S_i^{auto}} w(u,v_i^{auto})$, $v_i^{auto}$ is activated in time step 1; given $v_i^{auto}$ is activated, the probability that it turns positive is $1- x^{\top}_-(v_i^{auto})\beta^*$.

Suppose the algorithm has run $k$ epochs. Let $Z_1,\ldots, Z_{d'}$ be the number of times that node $v_i^{auto}$ is activated in time step $1$. Then we have $Z_i \sim Bin(k,w_i^{auto})$. Setting $\varepsilon_k = \sqrt{\frac{q \log k }{2k}}$, by Hoeffding's inequality, we have 
$$
\P \Big ( Z_i \geq k\cdot w_i^{auto} - \varepsilon_k \Big ) \geq 1- \exp\Big(- 2\varepsilon_k^2 k \Big) = 1- 1/k^{q}.
$$
By union of probability, we have 
$$
\P \Big (Z_i \geq k\cdot w_i^{auto} -\varepsilon_k,i=1,\ldots, d' \Big ) \geq 1- d'/k^{q}.
$$
Define $Z_k^{auto} = \sum_{i=1}^{d'}  (k\cdot w_i^{auto} - \varepsilon_k)\cdot x_{-}(v_i^{auto})x_{-}(v_i^{auto})^\top $, it is easy to see $\V_k \succeq Z_k^{auto}$ with probability at least $1-d'/k^{q}$. Then we conclude with probability at least $1-d'/k^{q}$, \begin{align*}
\lambda_{\min}(\V_k) \geq & \lambda_{\min}(Z_k^{auto}) \geq \min_{i=1}^{d'}  (k\cdot w_i^{auto} - \varepsilon_k) \cdot \lambda_{\min}\Big (  \sum_{i=1}^{d'} x_{-}(v_i^{auto})x_{-}(v_i^{auto})^\top  \Big ) \\
\geq & \min_{i=1}^{d'}  (k\cdot w_i^{auto} - \varepsilon_k)   \cdot \lambda_{\min}^{auto} =  \big ( k\cdot   w_{\min}^{auto}     - \varepsilon_k \big )  \cdot \lambda_{\min}^{auto},
\end{align*}
where the last inequality comes from Assumption  \ref{assump:exploration_nodes-2} that $\lambda_{\min} (  \sum_{i=1}^{d'} x_{-}(v_i^{auto})x_{-}(v_i^{auto})^\top   )  = \lambda_{\min}^{auto}$.

\QED



\section{Other Auxiliary Results}
\subsection{Lemma \ref{lemma:continuity} and Its Proof}\label{apd:aux-lemmas-continuity}
\begin{lemma}\label{lemma:continuity}
For any set of cardinality $K$ and any edge weights $p$, let $S^{\text{opt}}(p,\beta)$ be the optimal size-$K$ seed set. Then the optimal expected reward $f(S^{\text{opt}}(p,\beta), p,\beta)$ is continuous in both $p$ and $\beta$. 
\end{lemma}
\proof{Proof.}
First of all, for any fix seed set $S$, it is easy to see $f(S,p,\beta)$ is a continuous function with respect to $p$ and $\beta$. Let $S^{\text{opt}}(p_1,\beta_1)$ and $S^{\text{opt}}(p_2,\beta_2)$ be the optimal size-$K$ seed sets corresponding to pairs $(p_1,\beta_1)$ and $(p_2,\beta_2)$ respectively. For $(p_1,\beta_1)$, $(p_1,\beta_1)$  close enough, if $S^{\text{opt}}(p_1,\beta_1) = S^{\text{opt}}(p_1,\beta_1)$, then  $|f(S^{\text{opt}}(p_1,\beta_1), (p_1,\beta_1)) - f(S^{\text{opt}}(p_2), p_2)|$ can be bounded by a number small enough since the two functions share the same seed sets and are continuous in the diffusion probability.

We use $\nu = (p;\beta)$ to denote a vector concatenated by $p$ and $\beta$. 
For $\delta > 0 $ and $\nu_1,\nu_2$ such that $\| \nu_1 -\nu_2 \| \leq \delta$, we consider the case where $S^{\text{opt}}(\nu_1) \neq S^{\text{opt}}(\nu_2)$.
It is clear that $f(S^{\text{opt}}(\nu_1), \nu_2) \leq f(S^{\text{opt}}(\nu_2), \nu_2)$ and  $f(S^{\text{opt}}(\nu_2), \nu_1) \leq f(S^{\text{opt}}(\nu_1), \nu_1)$.
Without loss of generality, suppose $f(S^{\text{opt}}(\nu_1), \nu_1) \geq f(S^{\text{opt}}(\nu_2), \nu_2)$, then we obtain 
\begin{align*}
    f(S^{\text{opt}}(\nu_1), \nu_1) \geq f(S^{\text{opt}}(\nu_2), \nu_2)  \geq f(S^{\text{opt}}(\nu_1), \nu_2).
\end{align*}
Again, for fixed seed set $S^{\text{opt}}(\nu_1)$, by continuity of $f(S^{\text{opt}}(\nu_1),\nu)$ in $\nu$, there exists $\epsilon >0$ sufficient small such that 
$$
| f(S^{\text{opt}}(\nu_1), \nu_1) - f(S^{\text{opt}}(\nu_2), \nu_2)| \leq | f(S^{\text{opt}}(\nu_1), \nu_1) - f(S^{\text{opt}}(\nu_1), \nu_2)| \leq \epsilon.
$$
This way we conclude that $f(S^{\text{opt}}(\nu),\nu)$ is a continuous function with respect to $\nu$, which completes the proof.
\QED

\end{appendices}

\end{document}